\documentclass[prb,twocolumn,showpacs,amsmath,amssymb]{revtex4}

\usepackage{graphicx}
\usepackage{dcolumn}
\usepackage{bm}
\usepackage{color}

\usepackage{epsfig,float,afterpage,amssymb,wrapfig,psfrag}
\usepackage{subfigure}

\DeclareMathAlphabet{\bi}{OML}{cmm}{b}{it}

\newcommand{\kk}{\mathbf{k}}
\newcommand{\loc}{\mathrm{loc}}

\newcommand{\tbk}{\tilde{{\bf k}}}
\newcommand{\tbq}{\tilde{{\bf q}}}

\newcommand{\bQ}{{\bf Q}}

\begin{document}
\title{Cluster Extended Dynamical Mean Field Approach and Unconventional Superconductivity}

\author{J.\ H.\ Pixley\footnote{Present Address:
   Condensed Matter Theory Center and Joint Quantum Institute, Department of Physics,
University of Maryland, College Park, Maryland 20742-4111, USA } }
\affiliation{Department of Physics and Astronomy, Rice University,
Houston, Texas, 77005, USA}
\author{Ang Cai}
\affiliation{Department of Physics and Astronomy, Rice University,
Houston, Texas, 77005, USA}
\author{Qimiao Si}
\affiliation{Department of Physics and Astronomy, Rice University,
Houston, Texas, 77005, USA}

\date{\today}
\begin{abstract}
The extended dynamical mean field theory has played an important role in the study of quantum phase transitions in heavy fermion systems.  In order to incorporate the physics of unconventional superconductivity, we develop a cluster version of the extended dynamical mean field theory.  In this approach, we show how magnetic order and superconductivity develop as a result of inter-site spin exchange
interactions, and 
analyze in some detail the form of correlation functions.  We 
also
discuss the methods that can be used to solve the dynamical equations associated with this approach.  
Finally, we  
consider
different settings in which our approach can be applied, including the periodic Anderson model 
for heavy fermion systems.
\end{abstract}

\maketitle
\section{Introduction}

Unconventional superconductivity in heavy fermion metals often develops in the vicinity of antiferromagnetic (AF) order\cite{Mathur_Nature1998,Park_Nature2006,Stockert_Natphys2012}.
Its understanding is intimately connected with that of the  AF quantum critical points (QCPs) \cite{SiSteglich_Science2010}.
Traditional descriptions of heavy fermion quantum criticality are based on those for purely itinerant magnetism, 
in terms of the fluctuations of the spin-density-wave (SDW) order parameter \cite{Hertz,Millis,Moriya}. 
Studies in the recent past have emphasized the ``beyond Landau" physics of Kondo 
destruction \cite{Si-2001,Coleman-2001,Senthil-2004}.  In these studies,
considerable progress has been made based on 
the extended dynamical mean field theory (EDMFT) solution of Kondo lattice models \cite{Si_etal_JPSJ2013}.
One of the important questions along this direction concerns the implications of these theoretical studies on
 the understanding of superconductivity.

The EDMFT approach builds on the dynamical mean field theory (DMFT)~\cite{DMFT-review}, 
which was developed through the infinite dimensional limit of the Hubbard model~\cite{MetznerVollhardt}. 
DMFT maps an interacting lattice problem to a single quantum impurity model coupled to a self consistent Weiss field.
The latter captures the environment as seen at the single-particle level, and is represented by a fermionic bath.
DMFT has made significant contributions to a variety of strongly correlated problems~\cite{DMFT-review}, 
and has in particular provided significant new insights on the Mott transition.

The EDMFT approach  treats inter-site density-density
or spin-spin interactions, leading to a single impurity model coupled to self consistent fermionic and bosonic
baths\cite{Si-1996,Smith,Chitra-2000}.  
It has been extensively applied to the study of
AF quantum critical heavy fermion metals~\cite{Si-2003,Zhu.2003,Glossop.2007,Zhu.2007,Si_etal_JPSJ2013}.
For a Kondo lattice, the EDMFT approach yields
a Bose-Fermi Kondo model (BFKM) with self consistent bosonic and fermionic baths.
Kondo destruction arises from this approach.
This may be seen already at the level of the BFKM in the absence of self consistency.
The coupling of the local moment to the bosonic bath competes with the Kondo effect, i.e.
the tendency of singlet formation due to the
AF
 exchange coupling between the local moment and the fermionic bath.
 When the spectrum of the bosonic bath is sufficiently soft, corresponding to the spectral function being ``sub-ohmic",
this competition gives rise to the destruction of the Kondo effect, in a way that is associated with the criticality of the BFKM~\cite{Zhu-02}.
Studying this type of criticality in a variety of quantum impurity models has led to a number of new insights 
regarding Kondo destruction QCPs.
For the lattice case, through the EDMFT equations,
the bosonic spectrum is particularly soft near the 
AF QCP due to the critical slowing down, and one consistent 
solution is that the Kondo destruction occurs at the AF QCP.
There is by now extensive experimental evidence for this type
of beyond-Landau QCP from experiments in heavy fermion
metals, both in terms of 
a unusual scaling of the dynamical spin susceptibility in the quantum
critical regime and a sudden jump
of the Fermi surface across the QCP~\cite{Park_Nature2006,Aronson.95+Schroeder.00,Paschen.00+Gegenwart.07+Friedemann.10, Shishido, SiPashchen13}.
However, in order to study the important problem of the interplay between
this unconventional quantum criticality and superconductivity, 
a cluster generalization of the EDMFT is called for.
In this manuscript, we develop 
such a formalism.

In DMFT based approaches, incorporating real space correlations beyond a single site have naturally been done with the development of quantum cluster theories~\cite{cluster-review}.  In this case, strongly correlated problems can be mapped to a quantum cluster model 
with self consistent fermionic baths and the interactions within the cluster are treated exactly.  
Importantly,
dynamical cluster theories
incorporate non-perturbative corrections to DMFT without introducing a non-causal self energy~\cite{cluster-review}. 
This can be formulated in real space which leads to cluster DMFT (CDMFT)~\cite{cdmft},
or in momentum space which is known as the dynamical cluster approximation (DCA)\cite{DCA};
there are 
other cluster embedding schemes possible such as the variational cluster approximation (VCA)\cite{VCA}.
When the Weiss fields are neglected these cluster schemes are no longer self consistent and reduce to cluster perturbation theory (CPT),
which approximates lattice quantities by expanding about the isolated cluster limit~\cite{CPT}.   
A main advantage of dynamical cluster theories is the ability to account for 
various types of order
not possible within DMFT~\cite{Haule-2007}. For example, a four site cluster can treat a d-wave superconducting order parameter as well 
as stripe charge or spin order. Such a pairing mechanism is expected to be appropriate for, {\it e.g.}, the cuprates and heavy fermion materials.
 In this case, any superconducting ground state will have cooper pairs formed between sites which can lead to a variety of different pairing symmetries, such as extended s-wave, p-wave or d-wave. 

In this manuscript, we present a cluster extended dynamical mean field theory scheme that we dub C-EDMFT.  We derive the equations by generalizing ref.~\onlinecite{Smith} to the cluster case using a locator expansion about a dressed cluster limit.  
We formulate the equations in both real and momentum space.  We introduce magnetic order in the same fashion as EDMFT 
in refs.~\onlinecite{Zhu.2003,Glossop.2007,Zhu.2007} distinct from DMFT, and then generalize this approach 
to include superconductivity as well.  
We also construct the pairing correlation functions induced by magnetic interactions in the normal state within this approach.  
Lastly, we use the formalism to derive effective impurity models associated with strongly correlated problems of central interest.

We note that cluster generalizations of the EDMFT have been carried out in various forms in the past~\cite{DCA2,RVB-DMFT,Sun}. 
Where there is overlap, our approach is consistent with these formulations. 
We will make the specific comparisons as we go through the derivation of our approach. 
In short, the C-EDMFT formalism developed here brings out two new aspects (for definiteness, we will describe these with Eq.~(\ref{eqn:ham}) in mind). First, it is the inter-site $J_{ij}$ interactions which underlie both the magnetic and superconducting orders. Such orders develop through decoupling the $J_{ij}$ interaction term into the appropriate channels. Second, the approach avoids double-counting the inter-site interaction by suppressing the induced inter-site interactions associated with the polarization of the fermonic bath by the order parameter. As a result, the ${\bf q}$-dependence of the dynamical spin susceptibility arises through the $J_{ij}$ interaction term, instead of the bare particle-hole bubble at the ``special-${\bf q}$" (as opposed to the ``generic-${\bf q}$", see section III). These two aspects are in the same spirit as discussed for the case of the EDMFT \cite{Si-2005}.
  
\subsection{Development of Cluster EDMFT}

For illustration purposes, we consider a one band Hubbard model with two body inter-site interactions on a generic lattice.
\begin{eqnarray}
H&=& \sum_{\langle i,j \rangle, \sigma} t_{ij}(c_{i\sigma}^{\dag}c_{j\sigma} + \mathrm{h.c}) + U\sum_i n_{i\uparrow}n_{i\downarrow}
\nonumber
\\
&+& \sum_{\langle i,j \rangle,\alpha} J^{\alpha}_{ij} S^{\alpha}_i S^{\alpha}_j
\label{eqn:ham}
\end{eqnarray}
where $c_{i\sigma}$ destroys an electron of spin $\sigma$ at site $i$, $n_i=\sum_{\sigma} n_{i\sigma}$, 
and $n_{i\sigma}=c_{i\sigma}^{\dag}c_{i\sigma}$.  The index $\alpha$ runs over $0,1,2,3$, where 
for $\alpha=1,2,3$ (or for $\alpha=x,y,z$) the operator $S^{\alpha}_i \equiv c^{\dag}_{i\mu}(\sigma^{\alpha}_{\mu\nu}/2)c_{i\nu}$, 
is the spin operator, where $\sigma^{\alpha}_{\mu\nu}$ is the $\alpha$-Pauli matrix for $\alpha=x,y,z$.  
In addition we consider the charge channel with $\alpha=0$,  where the operator $S^{0} _i\equiv:n_{i}:$ denotes the 
normal ordered density $:n_i: = n_i - \langle n_i \rangle $.  We denote nearest neighbors by $\langle i,j \rangle$ and only 
consider nearest neighbor hopping $t_{ij}$ and two body exchange interaction $J^{\alpha}_{ij}$.  
For $J^{\alpha}_{ij}=0$, the model reduces to the standard Hubbard model with an onsite Coulomb repulsion of strength $U$.  
It is natural to extend these techniques to multi-band models and longer range interactions.  

A main focus of this work is a self consistent solution of the single particle Greens function 
$G_{ij\sigma}(\tau)=-\langle T_{\tau} c_{i\sigma}(\tau) c_{j\sigma}^{\dag} \rangle$ as well as the spin and charge susceptibilities 
$\chi_{ij}^{\alpha}(\tau)=\langle T_{\tau} :S^{\alpha}_i(\tau) ::S^{\alpha}_j : \rangle $.  In general, for the single particle Greens function, 
a perturbative expansion about the non-interacting limit yields the Dyson equation 
\begin{equation}
G({\bf k},i\omega_n) = \frac{1}{i\omega_n - \mu - t_{\bf k} - \Sigma_{\mathrm{lat}}({\bf k}, i\omega_n)},
\label{eqn:G}
\end{equation}
where $t_{\bf k}$ is the Fourier transform of $t_{ij}$$=1/N\sum_{\kk}e^{i\kk\cdot({\bf r}_i-{\bf r}_j)}t_{\kk}$, $\mu$ 
is the chemical potential, $\Sigma_{\mathrm{lat}}({\bf k}, i\omega_n)$ is the single particle self energy, 
and we denote fermionic Matsubara frequencies as $\omega_n$.
Analogous to the single particle Greens function, we introduce a spin and charge self energy $M_{\mathrm{lat}}^{\alpha}({\bf q},i\nu)$ 
which is defined in terms of each susceptibility as
\begin{equation}
\chi^{\alpha}({\bf q},i\nu_n) = \frac{1}{M^{\alpha}_{\mathrm{lat}}({\bf q},i\nu_n)+J_{{\bf q}}^{\alpha}}
\label{eqn:chi}
\end{equation}
where $J_{{\bf q}}^{\alpha}$ is the Fourier transform of $J^{\alpha}_{ij}$ and $\nu_n$ is a bosonic Matsubara frequency.  
The spin/charge self energies specify how much their corresponding susceptibilities differ from a Gaussian model where 
$\chi^{\alpha}_{ij} \propto 1/J^{\alpha}_{ij}$( ref.~\onlinecite{RVB-DMFT}).  In the following we will derive a self consistent 
C-EDMFT approach to approximate the lattice quantities $\Sigma_{\mathrm{lat}}({\bf k}, i\omega_n)$ and $M_{\mathrm{lat}}^{\alpha}({\bf q},i\nu)$ 
and in turn the single particle Greens function and spin/charge susceptibilities.

The remainder of the paper is organized as follows.
We focus on the C-EDMFT equations in the absence of any order in section \ref{sec:ECT},
and those in the presence of magnetic order in section \ref{sec:mag-order}.
We then apply the approach
to superconducting order and correlations in section \ref{sec:Sup}.  
We use the formalism to derive an effective cluster model in section \ref{sec:ECM}.
Finally, we outline 
the relevant solution methods in section \ref{sec:Sol_methods},
discuss several pertinent points in section \ref{sec:Dis},
before concluding  the paper in section \ref{sec:Con}.

\section{Normal State in the absence of Broken Symmetry}
\label{sec:ECT}
\begin{figure}[h!]
\begin{center}
\includegraphics[width=2.25in]{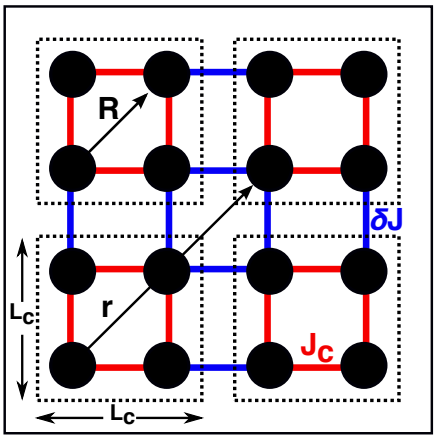}
\end{center}
\caption{(Color online) Division of the lattice into $N/N_c$ clusters, of size $N_c=L_c^{d_c}$, where $d_c$ is the dimensionality of the cluster.  The vector ${\bf r}$ labels each cluster, while sites within the cluster are labelled by ${\bf R}$.  Interactions are divided into within the cluster $J_c$ and between clusters $\delta J$, (which is also done for the hopping elements $t_c$ and $\delta t$.)   We have omitted the channel index $\alpha$ for clarity.
}
\label{fig:cluster}
\end{figure}

We begin by dividing the lattice of $N$ sites into clusters of size $N_c$, where each lattice site is now 
labeled by ${\bf x}={\bf r}+{\bf R}$, where ${\bf r}$ labels the cluster and ${\bf R}$ labels the sites within the cluster (see figure~\ref{fig:cluster}).  In the following we will use upper case latin letters to denote cluster indices (i.e. indices within the cluster) and lower case letters to label each cluster.  
This is then Fourier transformed to ${\bf k}=\tbk + {\bf K}$, where ${\bf K}$ is the intra cluster momentum and $\tbk$ 
 the inter cluster momentum.  With this notation $t_{ij}$ and $J^{\alpha}_{ij}$ can be written as 
 $A_{{\bf R}_i{\bf R}_j}({\bf r}_i - {\bf r}_j)={\bf A}({\bf r}_i - {\bf r}_j)$, where the bold ${\bf A}$ denotes a matrix in cluster indices. 
 We then separate $t_{ij}$ and $J^{\alpha}_{ij}$ into intra and inter cluster parts
\begin{eqnarray}
{\bf t}({\bf r}_i - {\bf r}_j) &=&  {\bf t}_c \delta_{{\bf r}_i ,{\bf r}_j} + \delta {\bf t}({\bf r}_i - {\bf r}_j),
\nonumber
\\
{\bf J}^{\alpha}({\bf r}_i - {\bf r}_j) &=&  {\bf J}_c^{\alpha} \delta_{{\bf r}_i ,{\bf r}_j} + \delta {\bf J}^{\alpha}({\bf r}_i - {\bf r}_j), 
\end{eqnarray}
where ${\bf t}_c$ and ${\bf J}_c^{\alpha}$ are the interactions within the cluster, whereas $\delta{\bf t}$ and $\delta{\bf J}^{\alpha}$ are the interactions between clusters, note that by construction $\delta{\bf t}(0)$ and $\delta{\bf J}^{\alpha}(0)$ vanish.
\subsection{Real Space Formulation} 
\label{subsec:ECTR}
We will first derive the equations in real space. We first focus on a ground state with no broken symmetry, and will then generalize 
the equations to the case of magnetic order and superconductivity in sections  \ref{sec:mag-order} and \ref{sec:Sup}.
 We perform a locator expansion in $\delta t$ and $\delta J$ about the cluster limit~\cite{cluster-review}.  
 The isolated cluster single particle Greens function and susceptibilities are defined by $C^0_G(X,Y;{\bf r}, \tau)= -\langle T_{\tau} 
 c_{{\bf r}X\sigma}(\tau) c_{{\bf r}Y\sigma}^{\dag} \rangle_{H_c}$ and $C^0_{\chi^{\alpha}}(X,Y;{\bf r}, \tau)= \langle T_{\tau} S^{\alpha}_{{\bf r}X\sigma}(\tau) S^{\alpha}_{{\bf r}Y\sigma} \rangle_{H_c}$ respectively, where $H_c$ is the \emph{isolated} cluster Hamiltonian 
 at cluster ${\bf r}$.  In the following we consider problems that have translational invariance between clusters which implies each 
 cluster correlation function is identical and we can drop the label ${\bf r}$.  
 
\begin{figure}[h!]
\begin{center}
\includegraphics[width=3.2in]{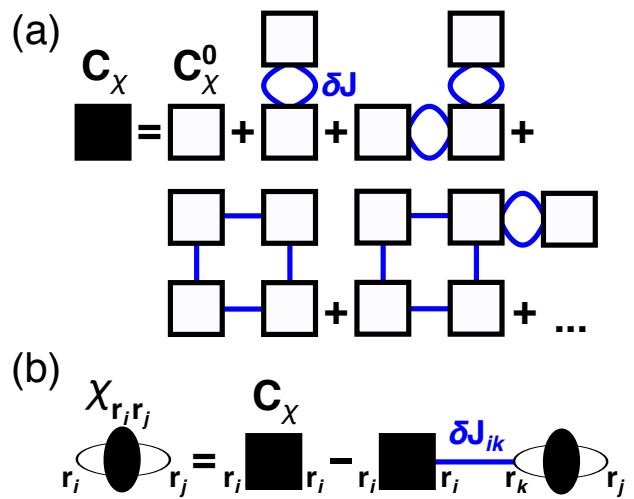}
\end{center}
\caption{(Color online) (a)  Real space diagrammatic representation of the effective cluster spin susceptibility $C_{\chi}$ (solid square), as an expansion in bare cumulants $C_{\chi}^0$ (empty square) with all local decorations that are completely irreducible by cutting a single $\delta J$ line.  (b)  Diagrammatic representation of the spin susceptibility in equation (\ref{eqn:c-chi}), expanding about the effective cluster limit.  Each term is for a specific two body interaction channel index $\alpha$, (with the index omitted for clarity).}
\label{fig:dyson}
\end{figure}

 We now generalize the effective cumulant expansion of Metzner~\cite{Metzner} for the Greens function and Smith and Si~\cite{Smith} for the susceptibilities from a single site to a cluster, which leads to matrix quantities.  Along these lines, we introduce the \emph{effective} cluster Greens function $C_G(X,Y; \tau)$ and spin/charge susceptibilities $C_{\chi^{\alpha}}(X,Y; \tau)$ which are defined as the isolated cluster Greens function and susceptibility (in the $\alpha$ channel) with all local decorations that are irreducible by cutting a single $\delta {\bf t}$ and $\delta {\bf J}^{\alpha}$ line respectively (see figure~\ref{fig:dyson}).  
 The effective cluster correlation functions can be regarded as ``dressed'' cluster correlation function, generalizing the dressed atom picture~\cite{Metzner} to the cluster case.  Retaining this class of diagrams can be formally justified in the large dimensional limit after rescaling  $\delta {\bf t}$ and $\delta {\bf J}^{\alpha}$ by the square root of the coordination raised to the manhattan distance between clusters while keeping the dimension and number of sites in the cluster fixed\cite{cdmft}.  
 Performing the locator expansion about the effective cluster correlation functions we arrive at the following Dyson like equations
\begin{eqnarray}
{\bf G}_{{\bf r}_i{\bf r}_j}(i \omega_n) &=& {\bf C}_G(i\omega_n)\delta_{{\bf r}_i,{\bf r}_j}
\nonumber
\\&+&{\bf C}_G(i\omega_n)\sum_{{\bf r}_l}\delta{\bf t}({\bf r}_l-{\bf r}_j){\bf G}_{{\bf r}_l{\bf r}_j}(i \omega_n)
\label{eqn:c-G}
\\
\bm{\chi}^{\alpha}_{{\bf r}_i{\bf r}_j}(i \nu_n) &=& {\bf C}_{\chi^{\alpha}}(i\nu_n)\delta_{{\bf r}_i,{\bf r}_j}
\nonumber
\\&-&{\bf C}_{\chi^{\alpha}}(i\nu_n)\sum_{{\bf r}_l}\delta{\bf J}^{\alpha}({\bf r}_l-{\bf r}_j)\bm{\chi}^{\alpha}_{{\bf r}_l{\bf r}_j}(i \nu_n).
\label{eqn:c-chi}
\end{eqnarray}
We note that these are matrix equations and we are using a bold notation to denote matrices in cluster indices.  Fourier transforming equations (\ref{eqn:c-G}) and (\ref{eqn:c-chi}) to inter-cluster momentum $\tbk$ and $\tbq$ we arrive at for the Greens function
\begin{eqnarray}
{\bf G}(\tbk,i \omega_n) &=& {\bf C}_G(i\omega_n)
+{\bf C}_G(i\omega_n)\delta{\bf t}(\tbk){\bf G}(\tbk,i \omega_n)
\\
&=& \left[{\bf C}_G(i\omega_n)^{-1} -\delta{\bf t}(\tbk) \right]^{-1}
\end{eqnarray}
and for the susceptibility
\begin{eqnarray}
\bm{\chi}^{\alpha}(\tbq,i \nu_n) &=& {\bf C}_{\chi^{\alpha}}(i\nu_n)
-{\bf C}_{\chi^{\alpha}}(i\nu_n)\delta{\bf J}^{\alpha}_{\tbq}\bm{\chi}^{\alpha}(\tbq,i \nu_n)
\\
&=&\left[{\bf C}_{\chi^{\alpha}}(i\nu_n)^{-1} + \delta {\bf J}^{\alpha}_{\tbq} \right]^{-1}.
\end{eqnarray}
Rewriting the Dyson equations for $G$ and $\chi^{\alpha}$ in equations (\ref{eqn:G}) and (\ref{eqn:chi}) in real space cluster indices we find the self energy and spin/charge self energies are $\tbk$ and $\tbq$ independent respectively, and only depend on cluster indices.  We arrive at the following equations for the effective cluster correlation functions 
\begin{eqnarray}
{\bf C}_G(i\omega_n)^{-1} &=&{\bf g}^{0}_c(i\omega_n)^{-1} - \bm{\Sigma}(i\omega_n)
\\
{\bf C}_{\chi^{\alpha}}(i\nu_n)^{-1} &=&  {\bf J}_c^{\alpha}+{\bf M}^{\alpha}(i\nu_n),
\end{eqnarray}
where the free isolated cluster Greens function is ${\bf g}^{0}_c(i\omega_n)^{-1}=(i\omega_n + \mu)\bm{1} - {\bf t}_c$ and $\bm{1}$ is the identity matrix in cluster indices.

The fact that both self energies are $\tbk$ and $\tbq$ independent implies they can be calculated by an effective cluster model.  The effective cluster model can be obtained through a generalized cavity method~\cite{cdmft}, by expanding the partition function in terms of $\delta {\bf t}$ and $\delta {\bf J}^{\alpha}$ about a particular cluster ${\bf o}$ and effectively integrating out all other degrees of freedom. This leads to the cluster action
\begin{eqnarray}
&S_C&= S_C^0 - \int_0^{\beta}d\tau d\tau^{\prime} \sum_{X,Y,\sigma} c_{X\sigma}^{\dag}(\tau)\mathcal{G}_{0,XY}^{-1}(\tau-\tau^{\prime})c_{Y\sigma}(\tau^{\prime})
\nonumber
\\
&-&\frac{1}{2}\int_0^{\beta}d\tau d\tau^{\prime} \sum_{X,Y,\alpha} S^{\alpha}_X(\tau)\chi_{0\alpha,XY}^{-1}(\tau-\tau^{\prime})S^{\alpha}_Y(\tau^{\prime}).
\label{eqn:c-action}
\end{eqnarray}
We have dropped the cluster label ${\bf o}$, defined the isolated cluster action as 
\begin{eqnarray}
S_C^0&=&\int_0^{\beta}d\tau \,U\sum_{X\in C} n_{X\uparrow}(\tau)n_{X\downarrow}(\tau) 
\nonumber
\\
&+& \sum_{\langle X,Y \rangle,\alpha}J_{c,XY}^{\alpha}S^{\alpha}_X(\tau)S^{\alpha}_Y(\tau),
\label{eqn:isolated-action}
\end{eqnarray}
 and introduced the effective Weiss fields $\mathcal{G}_{0,XY}^{-1}$ and $\chi_{0\alpha,XY}^{-1}$ that are related to 
lattice quantities by
\begin{eqnarray}
\bm{\mathcal{G}}_0^{-1}(i\omega_n) &=& {\bf g}_c(i\omega_n)^{-1} - \sum_{{\bf r}_i,{\bf r}_j}\delta {\bf t}_{{\bf o}{\bf r}_i}{\bf G}^{({\bf o})}_{{\bf r}_i{\bf r}_j}(i\omega_n)\delta{\bf t}_{{\bf r}_j{\bf o}}
\,\,\,\,\,\,\,\,\,\,\,
\\
\bm{\chi}_{0\alpha}^{-1}(i\nu_n) &=& \sum_{{\bf r}_i,{\bf r}_j}\delta {\bf J}^{\alpha}_{{\bf o}{\bf r}_i}\bm{\chi}^{\alpha({\bf o})}_{{\bf r}_i{\bf r}_j}(i\nu_n)\delta{\bf J}^{\alpha}_{{\bf r}_j{\bf o}}.
\label{eqn:cav-chi}
\end{eqnarray}
Where ${\bf G}^{({\bf o})}$ and $\bm{\chi}^{\alpha({\bf o})}$ are the Greens function and spin susceptibility of the lattice with the cluster ${\bf o}$ removed, and we have taken the cluster ${\bf o}$ to be at the origin.
Generalizing the arguments of ref.~\onlinecite{Smith} to the case of matrix cluster quantities we can relate ${\bf G}^{({\bf o})}$ and $\bm{\chi}^{\alpha({\bf o})}$ to the full Greens function and spin/charge susceptibilities to obtain (omitting the frequency labels)
\begin{eqnarray}
{\bf G}^{({\bf o})}_{{\bf r}_i{\bf r}_j} &=& {\bf G}_{{\bf r}_i{\bf r}_j} - {\bf G}_{{\bf r}_i{\bf o}}({\bf G}_{{\bf o}{\bf o}})^{-1}
{\bf G}_{{\bf o}{\bf r}_j}
\\
\bm{\chi}^{\alpha({\bf o})}_{{\bf r}_i{\bf r}_j} &=&\bm{\chi}^{\alpha}_{{\bf r}_i{\bf r}_j} - \bm{\chi}^{\alpha}_{{\bf r}_i{\bf o}}(\bm{\chi}^{\alpha}_{{\bf o}{\bf o}})^{-1}
\bm{\chi}^{\alpha}_{{\bf o}{\bf r}_j}.
\end{eqnarray}
With these relations, equations (\ref{eqn:c-G}) and (\ref{eqn:c-chi}), as well as the self consistency conditions
\begin{eqnarray}
{\bf G}_{\mathrm{loc}}(i\omega_n) &=& \frac{N_c}{N}\sum_{\tbk}{\bf G}(\tbk,i\omega_n),
\label{eqn:self-G}
\\
\bm{\chi}^{\alpha}_{\loc}(i\nu_n) &=& \frac{N_c}{N}\sum_{\tbq}\bm{\chi}^{\alpha}(\tbq,i\nu_n),
\label{eqn:self-chi}
\end{eqnarray}
 the Weiss fields are completely determined by the self energies and the local correlation functions where
\begin{eqnarray}
\bm{\mathcal{G}}_0^{-1}(i\omega_n) &=& \bm{\Sigma}(i\omega_n) + {\bf G}_{\mathrm{loc}}(i\omega_n)^{-1},
\\
\bm{\chi}_{0\alpha}^{-1}(i\nu_n) &=& {\bf M}^{\alpha}(i\nu_n) + {\bf J}^{\alpha}_c - \bm{\chi}^{\alpha}_{\mathrm{loc}}(i\nu_n)^{-1}.
\end{eqnarray}
We have used equations (\ref{eqn:c-G}) and (\ref{eqn:c-chi}) to eliminate the dependence on the effective cumulants 
and use the subscript ``$\loc$'' denote averages calculated with the effective cluster action in equation (\ref{eqn:c-action}), 
which also corresponds to the lattice quantities within the cluster as enforced via the self consistent equations (\ref{eqn:self-G}) 
and (\ref{eqn:self-chi}).  
One can generalize the arguments of refs.~\onlinecite{cdmft,DCA} to prove that the approach here 
has manifestly causal self energies for both $\bm{\Sigma}$ and ${\bf M}^{\alpha}$.

It is useful to consider a few limiting cases of the above equations.  First, we note that setting 
$\bm{\mathcal{G}}_0^{-1}(i\omega_n)={\bf g}^{0}_c(i\omega_n)^{-1}$ and $\bm{\chi}_{0\alpha}^{-1}(i\nu_n)=0$
 implies the effective cluster correlation functions reduce to the isolated cluster quantities, and the self energies are then 
 completely determined by solving the isolated cluster problem.  
 Therefore, in the absence of Weiss fields this approach reduces to CPT for both $G_{ij}$ 
 and $\chi_{ij}^{\alpha}$ and is no longer self consistent.  This clarifies the meaning of keeping all local decorations for $C_G$
  and $C_{\chi^{\alpha}}$ and is necessary to properly introduce the dynamical Weiss fields.  
  In the limit of one site in the cluster $N_c =1$, the equations reduce to EDMFT; this underscores the fact that
   the cluster theories incorporate spatial fluctuations beyond standard dynamical mean field theories.  
   Lastly, in the limit of large cluster sizes, $N_c\rightarrow\infty$ the theory becomes exact. 
    In this sense, extended dynamical cluster theories interpolates between the EDMFT and the exact answer as the cluster size is increased.

After self consistency has been reached it is possible to restore translational invariance to the self energies and thereby the correlation functions by interpolating the cluster quantities.  Since the self energies are only defined for sites within the cluster (or cluster momentum) the interpolation scheme must respect the symmetry of the original lattice.  Following ref.~\onlinecite{cluster-review}, after the self consistent solution has been reached we interpolate the cluster self energies to obtain the lattice quantities with the estimation
\begin{eqnarray}
\Sigma_{\mathrm{lat}}({\bf k},i\omega_n) &=& \frac{1}{N_c}\sum_{{\bf X},{\bf Y}}e^{-i{\bf k}\cdot({\bf X}-{\bf Y})}\Sigma({\bf X},{\bf Y},i\omega_n), \,\,\,\,\,\,\,\,\,\,\,\,\,
\\
M^{\alpha}_{\mathrm{lat}} ({\bf q},i\nu_n)&=&\frac{1}{N_c}\sum_{{\bf X},{\bf Y}}e^{-i{\bf q}\cdot({\bf X}-{\bf Y})}M^{\alpha}({\bf X},{\bf Y},i\nu_n).
\end{eqnarray}
We then use these to determine the lattice Greens function and spin susceptibility in equations (\ref{eqn:G}) and (\ref{eqn:chi}).  Other interpolation schemes are possible as described in ref.~\onlinecite{biroli2002}, where each scheme preserves the symmetry of the lattice.  

Our equations without restoring the translational invariance have some similarities with those of ref.~\onlinecite{Sun}, which invoked a different procedure. Namely, a Hubbard-Stratonovich transformation decouples the inter site two body interaction term, and the Hubbard-Stratonovich field becomes a self consistent dynamic bosonic Weiss field in the cluster limit. We also note that, in the absence of conduction electrons similar equations for the self consistent bosonic bath appeared in the construction for spin only models in ref.~\onlinecite{RVB-DMFT}.

\subsection{Momentum Space Formulation} 
\label{subsec:ECTM}
The momentum space construction parallels the DCA formulation, which restores translation symmetry by giving the cluster periodic boundary conditions\cite{DCA}.  This is achieved by modifying the cluster Fourier transform to 
\begin{eqnarray}
[J^{\alpha}_{\mathrm{DCA}}(\tilde{{\bf q}})]_{{\bf X}_i,{\bf X}_j}&\equiv&J^{\alpha}(\tilde{{\bf q}})_{{\bf X}_i,{\bf X}_j}e^{-i\tilde{{\bf q}}\cdot({\bf X}_i-{\bf X}_j)}
\label{eqn:dcaft}
\\
&=&\frac{1}{N_c}\sum_{{\bf Q}}e^{i{\bf Q}\cdot({\bf X}_i-{\bf X}_j)}J^{\alpha}_{{\bf Q}+\tilde{{\bf q}}}
\end{eqnarray}
In the following we will refer to this as the DCA Fourier transform.  This leads to periodic boundary conditions in the cluster and a coarse graining of the cluster quantities $\bar{t}_{{\bf K}}=N_c/N\sum_{\tbk}t_{\tbk +{\bf K}}$ and  $\bar{J}^{\alpha}_{{\bf Q}}=N_c/N\sum_{\tbq}J^{\alpha}_{\tbq +{\bf Q}}$ and in turn, the inter-cluster quantities become $\delta t_{\tbk +{\bf K}}=t_{\tbk +{\bf K}}-\bar{t}_{{\bf K}}$ and $\delta J^{\alpha}_{\tbq +{\bf Q}}=J^{\alpha}_{\tbq +{\bf Q}}-\bar{J}^{\alpha}_{{\bf Q}}$.  Applying the DCA Fourier transform to equations (\ref{eqn:c-G}) and (\ref{eqn:c-chi}) leads to matrix equations that are diagonal in momentum space.  Here we specify the equations for the spin susceptibility, with the equations for the Greens function being identical in form to those of DCA.  We have a Dyson like equation
\begin{eqnarray}
\chi^{\alpha}(\tbq+{\bf Q},i \nu_n) =\frac{1}{1/C_{\chi^{\alpha}}({\bf Q},i\nu_n) + \delta J^{\alpha}_{\tbq+{\bf Q} }},
\label{eqn:chi-q}
\end{eqnarray}
with an effective spin cumulant
\begin{equation}
1/C_{\chi^{\alpha}}({\bf Q},i\nu_n) = M^{\alpha}({\bf Q},i\nu_n) + \bar{J}^{\alpha}_{{\bf Q}} .
\end{equation}
The Weiss field is specified by
\begin{equation}
\chi_{0\alpha}^{-1}({\bf Q},i\nu_n) = M^{\alpha}({\bf Q},i\nu_n) + \bar{J}^{\alpha}_{{\bf Q}} - 1/\chi^{\alpha}_{\loc}({\bf Q},i\nu_n),
\end{equation}
with the self consistent equation 
\begin{equation}
\chi_{\loc}^{\alpha}({\bf Q},i\nu_n) = \frac{N_c}{N}\sum_{\tbq}\chi^{\alpha}(\tbq+{\bf Q},i\nu_n).
\label{eqn:self-chi-Q}
\end{equation}
Lastly, the cluster action is now diagonal in cluster momentum, which leads to
\begin{eqnarray}
S_C&=& S_C^0 - \int_0^{\beta}d\tau d\tau^{\prime} \sum_{{\bf K},\sigma} c_{{\bf K}\sigma}^{\dag}(\tau)\mathcal{G}_{0,{\bf K}}^{-1}(\tau-\tau^{\prime})c_{{\bf K}\sigma}(\tau^{\prime})
\nonumber
\\
&-&\frac{1}{2}\int_0^{\beta}d\tau d\tau^{\prime} \sum_{{\bf Q},\alpha} S^{\alpha}_{{\bf Q}}(\tau)\chi_{0\alpha,{\bf Q}}^{-1}(\tau-\tau^{\prime})S^{\alpha}_{-{\bf Q}}(\tau^{\prime}),\,\,\,\,\,
\label{eqn:c-action2}
\end{eqnarray}
where the isolated cluster action  [in equation (\ref{eqn:isolated-action})] is written in cluster momentum 
$S_C^0=\int_0^{\beta}d\tau \,[U\sum_{{\bf Q}} n_{{\bf Q}\uparrow}(\tau)n_{-{\bf Q}\downarrow}(\tau) + \sum_{{\bf Q},\alpha}\bar{J}^{\alpha}_{{\bf Q}}S^{\alpha}_{{\bf Q}}(\tau)S^{\alpha}_{-{\bf Q}}(\tau)]$. 
These equations are similar to those of ref.~\onlinecite{DCA2}, where the EDMFT approach was adopted through a Hubbard-Stratonovich transformation that decouples the inter-site interaction terms, and then the DCA approach is applied.

\subsection{Momentum dependence of the lattice susceptibility}
\label{subsec:q-dependence}
In both the real space and momentum space formulations, it is seen that the momentum dependence of the dynamical lattice spin susceptibility reflects the momentum dependence associated with the inter-site interaction $J_{ij}$. 
This feature is similar to what happens in
the EDMFT \cite{Si-1996,Smith,Chitra-2000,Si-2005},
with the advantage that the inter-site interactions that give rise to the momentum dependence of the dynamical susceptibility also appear in the self-consistent dynamical equations.

It is useful to stress that the momentum dependence 
of the lattice  susceptibility in the paramagnetic phase 
does not reflect that of the bare particle-hole bubble \cite{Si-2005}.
This is to be contrasted with the standard DMFT and cluster generalizations, where the bare particle-hole bubble
is responsible for the momentum dependence of the lattice susceptibility, which we discuss in detail in the following section.

\section{Magnetic Order}
\label{sec:mag-order}
Before we discuss the cluster generalization of symmetry broken phases within EDMFT, 
we find it very useful to review the different schemes used to introduce magnetic order in the single site case.
 In the context of the EDMFT, there are two ways to introduce magnetic order into the system, namely whether or not the magnetic order parameter polarizes the single particle Weiss field $\mathcal{G}_{0,\sigma}^{-1}$ (see ref.~\onlinecite{Si-2005} for details).  
 It has been shown~\cite{Si-2005} that allowing $\mathcal{G}_{0,\sigma}^{-1}$ to polarize amounts to keeping the particle hole bubble contribution, $[\chi_{\mathrm{ph}}({\bf q},\omega)]^{-1}-[\chi_{\mathrm{ph},\loc}(\omega)]^{-1}$, 
 to the spin susceptibility~\cite{DMFT-review} (where the particle hole bubbles are constructed using the full lattice and local 
 Greens function respectively obtained with DMFT and the brackets $[\dots ]$ denote a matrix form~\cite{Smith}).  
 Within the context of DMFT, such a term will exist due to the distinction between ``normal'' and ``special'' ${\bf q}$'s; {\it cf.} 
 ref.~\onlinecite{DMFT-review}.  However, due to promoting $J_{ij}$ to the same level as $t_{ij}$ within the EDMFT, 
 there are no special ${\bf q}$'s allowed, since this would make $J({\bf q}) \sim O(\sqrt{d})$ which 
 would diverge in the large $d$ limit.  
 The absence of any special ${\bf q}$'s
 implies $[\chi_{\mathrm{ph}}({\bf q},\omega)]^{-1}=[\chi_{\mathrm{ph},\loc}(\omega)]^{-1}$, and the particle hole bubble contribution vanishes~\cite{Smith}.  Keeping the particle hole bubble within the EDMFT amounts to double counting contributions from the spin-spin interaction~\cite{Si-2005}.  
 
 In the following section, we focus on the cluster EDMFT case.
This is to be contrasted with DCA, which parallels DMFT and therefore 
retains the distinction between special and generic $\tilde{\bf q}$'s and
a similar particle hole contribution to the spin susceptibility, namely $[\bm{\chi}_{\mathrm{ph}}(\tilde{{\bf q}},\omega)]^{-1}-[\bm{\chi}_{\mathrm{ph},\loc}(\omega)]^{-1}$  (here the particle hole bubble is constructed with the full single particle lattice and local cluster Greens function respectively, obtained within DCA and the bold denotes matrices in cluster momentum)~\cite{cluster-review}.  
As in EDMFT, the cluster EDMFT promotes $\delta J$ to the same level as $\delta t$. 
Generalizing the EDMFT argument~\cite{Smith} to the cluster case we conclude that there should only 
be generic $\tilde{{\bf q}}$'s.  This amounts to not allowing the single particle Weiss field $\bm{\mathcal{G}}_{0\sigma}^{-1} $
to polarize (i.e. is $\sigma$ independent), and is equivalent to the suppression of the particle hole bubble contribution.  
 
We now consider magnetic order with an ordering wave vector ${\bf q}={\bf q}_{\mathrm{or}}\equiv\tbq_{\mathrm{or}}+{\bf Q}_{\mathrm{or}}$ within the channel $\alpha=\lambda$ with $\lambda\neq 0$. The cluster chosen must be large enough 
to accommodate the type of magnetic order under consideration, for example a four site cluster can describe the collinear AF 
order with ${\bf q}_{\mathrm{or}}=(0,\pi)$ whereas a two site cluster can only treat either ferro- or
AF
order.  This then implies that the order pattern within each cluster must be the same and therefore $\tbq_{\mathrm{or}}=0$.

We will consider the equations in real space and momentum space consecutively.  
After separating $J_{ij}^{\alpha}$ into inter- and intra-cluster parts we treat the cluster interactions exactly and normal order the interaction between clusters, via $S^{\alpha}_i=:\hspace{-1.8mm}S^{\alpha}_i\hspace{-1.8mm}: + \langle S^{\alpha}_i \rangle$. 
All of the previous steps apply, but now we perform the locator expansion in the normal ordered interaction between clusters $\delta J^{\alpha}_{XY}({\bf r}_i-{\bf r}_j):S^{\alpha}_{{\bf r}_i X}: :S^{\alpha}_{{\bf r}_j Y}:$.  This corresponds to adding an additional term to the cluster action
\begin{equation}
S_C \rightarrow S_C -\int_0^{\beta}d\tau\,\sum_{X}h_{\loc}^{X}S^{\lambda}_{X}(\tau)
\end{equation}
and the local magnetic field is determined self consistently from
\begin{equation}
h_{\loc}^{X}=-\sum_{Y}\left[\delta J^{\lambda}_{XY}(\tbq_{\mathrm{or}}=0) + \chi_{0\lambda,XY}^{-1}(i\nu_n=0)\right]M_{Y},
\end{equation}
where $M_{Y}=\langle S^{\lambda}_{Y}\rangle_C$ and the average is over the cluster action.

In momentum space, this approach amounts to adding to the action
\begin{equation}
S_C \rightarrow S_C -\int_0^{\beta}d\tau\,h_{\loc}S^{\lambda}_{{\bf -Q}_{\mathrm{or}}}(\tau). 
\label{eqn:dcaordered}
\end{equation}
Now the local field is given by
\begin{equation}
h_{\loc} = -\left[\delta J^{\lambda}_{{\bf Q}_{\mathrm{or}}} + \chi_{0\lambda}^{-1}({\bf Q}_{\mathrm{or}},i\nu_n=0)\right]M
\label{eqn:hloc-Q}
\end{equation}
where $M=\langle S^{\lambda}_{{\bf Q}_{\mathrm{or}}}\rangle_C$ and in this case the inter-cluster interaction is $\delta J^{\lambda}_{{\bf Q}_{\mathrm{or}}}= J^{\lambda}_{{\bf Q}_{\mathrm{or}}}-\bar{J}^{\lambda}_{{\bf Q}_{\mathrm{or}}}$ where we have coarse grained the cluster interaction as described previously.

It is useful to note that, in the limit of no dynamical Weiss field, the mean field equations for the self consistent field $h_{\loc}$
reduce to that of cluster Weiss mean field theory.  Here, the dynamical Weiss field renormalizes the static field due to 
the dynamical interactions mediated by $\chi_{0\lambda}^{-1}$.

\subsection{Alternate Derivation of Spin Susceptibility}
\label{sec:LT}
In the following section, we use the self consistent equations that incorporate magnetic order to provide 
an alternative way of deriving the lattice spin susceptibility at an ordering wave vector ${\bf Q}_{\mathrm{or}}$, 
which we define as $\chi^{z}({\bf Q}_{\mathrm{or}},i\nu_{n})\equiv \chi_{or}(i\nu_{n})$.  
This approach will also provide insight into the way a superconducting long range order can be incorporated 
in C-EDMFT. In order to treat an ordering wave vector more naturally, we will consider the momentum space formalism.

In order to ease the notation in the following subsection we use a subscript to denote each Matsubara frequency, e.g. $S^{\alpha}_{i,n}(i\nu_{n})=S^{\alpha}_{i,n}$.
First we observe that if we include a dynamic source field for each Matsubara frequency $i\nu_n$, we must normal order every frequency modes instead of the static mode only, by writing 
$S^{\alpha}_{i,n}=:S^{\alpha}_{i,n}: + \langle S^{\alpha}_{i,n} \rangle$, as a result equation (\ref{eqn:dcaordered}) will become 
\begin{equation}
S_C \rightarrow S_C -\frac{1}{\beta}\sum_{n} h_{\loc,n} S^{\lambda}_{{\bf -Q}_{\mathrm{or}},-n},
 \label{eqn:newSC}
 \end{equation} 
with $h_{\loc,n} = -\left[\delta J^{\lambda}_{{\bf Q}_{\mathrm{or}}} + \chi_{0\lambda}^{-1}({\bf Q}_{\mathrm{or}},i\nu_n)\right] \langle S^{\lambda}_{{\bf Q}_{\mathrm{or}},n} \rangle
$.

We introduce an additional term, $S_{L}$, to the lattice action which couples the $\lambda=z$ 
component of the spin operators at wave vector $\pm{\bf Q}_{\mathrm{or}}$ to a dynamic source field
\begin{eqnarray}
S_{L}=-\frac{1}{\beta} \sum_{n} 
\left( h_{-{\bf Q}_{\mathrm{or}},-n} S^{z}_{{\bf q}={{\bf Q}_{\mathrm{or}}},n}
+ h_{{\bf Q}_{\mathrm{or}},n} S^{z}_{{\bf q}=-{{\bf Q}_{\mathrm{or}}},-n} \right).
\nonumber
\\
\end{eqnarray}
Mapping this into the cluster action, and assuming that the Weiss fields cannot be polarized 
(due to the absence of any special $\tbq$'s), 
we end up 
adding  \emph{only} one extra term $S_{1}$ to equation (\ref{eqn:dcaordered}), namely
\begin{equation}
S_{1}=-\frac{1}{\beta}  \sqrt{\frac{N_{c}}{N}} \sum_{n}  \left( h_{-{\bf Q}_{\mathrm{or}},-n} S_{{\bf Q}_{\mathrm{or}},n}^{z}+ h_{{\bf Q}_{\mathrm{or}},n} S_{-{\bf Q}_{\mathrm{or}},-n}^{z}\right)
\label{coarsegrained_h}
\end{equation}
Here $S_{{\bf Q}_{\mathrm{or}}}^{z}$, which is defined in the cluster, need to be distinguished from $S^{z}_{{\bf q}={{\bf Q}_{\mathrm{or}}}}$, which is defined on the lattice, since we have $S^{z}_{{\bf Q}_{\mathrm{or}}}=(1/\sqrt{N_{c}}) \sum_{\bf K} c_{\bf K+{\bf Q}_{\mathrm{or}}\alpha}^{\dagger} (\sigma^{z}_{\alpha\beta}/2)c_{{\bf K} \beta}$ and $S^{z}_{{\bf q}={{\bf Q}_{\mathrm{or}}}}=(1/\sqrt{N}) \sum_{\bf k} c_{\bf k+{\bf Q}_{\mathrm{or}} \alpha}^{\dagger} (\sigma^{z}_{\alpha\beta}/2) c_{{\bf k}\beta}$ respectively. 

The self consistency between the lattice and the cluster quantities implies the following.
\begin{equation}
\langle {S}_{{\bf Q}_{\mathrm{or}},n}^{z} \rangle _{C} = \sqrt{\frac{N_{c}}{N}} \langle S^{z}_{{\bf q}={\bf Q}_{\mathrm{or}},n} \rangle _{L},
 \label{eqn:self_con_for_S}
\end{equation}
Here $\langle \dots \rangle_C$ and $\langle \dots \rangle _{L}$ denote averaging over cluster action of equation (\ref{eqn:newSC}) and the lattice action with the additional source field term $S_{L}$ and $S_{1}$ respectively. 
To stress that the expectation value $\langle {S}_{{\bf Q}_{\mathrm{or}},n}^{z} \rangle _{C}$ is calculated self consistently we define $\langle {S}_{{\bf Q}_{\mathrm{or}},n}^{z} \rangle _{C} \equiv f_{n}(\langle {S}_{{\bf Q}_{\mathrm{or}},n}^{z}\rangle _{C}, h_{{\bf Q}_{\mathrm{or}},n},h_{-{\bf Q}_{\mathrm{or},-n}})$.

Differentiating both sides of equation (\ref{eqn:self_con_for_S}) with respect to $h_{{\bf Q}_{\mathrm{or}},n} $ at $h_{\pm {\bf Q}_{\mathrm{or}}, \pm n} =0$, the right-hand side gives the lattice correlation function,
\begin{eqnarray}
\frac{d \langle S^{z}({\bf q}={\bf Q}_{\mathrm{or}},i\nu_{n}) \rangle_{L}}{d h_{{\bf Q}_{\mathrm{or}},n} } &=& \frac{1}{\beta} \langle :S^{z}_{-{\bf Q}_{\mathrm{or}},-n}::S^{z}_{{\bf Q}_{\mathrm{or}},n}:\rangle_L 
\nonumber
\\
&=&\chi_{or}(i\nu_{n}) \label{eqn:dfdh}
\end{eqnarray}
 Whereas the left-hand side gives the cluster correlation function, using
 
 \begin{equation}
\frac{ d f_{n} } { d h_{{\bf Q}_{\mathrm{or}},n} }= \frac{ \partial f_{n}}{\partial \langle {S}_{{\bf Q}_{\mathrm{or}},n}^{z}\rangle _{C}}  \frac{ d \langle {S}_{{\bf Q}_{\mathrm{or}},n}^{z} \rangle _{C} } { d h_{{\bf Q}_{\mathrm{or}},n}}  + \frac {\partial f_{n} } {\partial h_{{\bf Q}_{\mathrm{or}},n} }
  \end{equation}
and,
\begin{eqnarray}
\frac{\partial f_{n}}{\partial h_{{\bf Q}_{\mathrm{or}},n} }&=&\sqrt{\frac{N_{c}}{N}} \frac{1}{\beta} \langle :S^{z}_{-{\bf Q}_{\mathrm{or}},-n} ::S^{z}_{{\bf Q}_{\mathrm{or}},n}:\rangle_{C}
\nonumber
\\
&=& \sqrt{\frac{N_{c}}{N}} \chi_{\mathrm{loc,or}} (i\nu_{n})
\label{eqn:pfph}
\\
\frac{\partial f_{n}}{\partial \langle {S}_{{\bf Q}_{\mathrm{or}},n}^{z} \rangle _{C} }&=&\frac{d h_{\loc,n}}{d \langle {S}_{{\bf Q}_{\mathrm{or}},n}^{z} \rangle _{C}  } \frac{\partial f_{n}}{\partial h_{\loc,n}} 
\\
&=&- \left[\delta J^{\lambda}_{{\bf Q}_{\mathrm{or}}} + \chi_{0\lambda}^{-1}({\bf Q}_{\mathrm{or}},i\nu_n)\right] \chi_{\mathrm{loc,or}}(i\nu_{n})
\nonumber
\label{eqn:pfpm}
\end{eqnarray}
where we have defined $\chi_{\mathrm{loc}}^{z}({\bf Q}_{\mathrm{or}},i\nu_{n}) \equiv  \chi_{\mathrm{loc,or}}(i\nu_{n})$ as the local spin susceptibility in the ordering channel $\bQ_{\mathrm{or}}$. We also have $d \langle {S}_{{\bf Q}_{\mathrm{or}},n}^{z} \rangle _{C} / d h_{{\bf Q}_{\mathrm{or}}, n}=\sqrt{{N_{c}}/{N}} \chi_{or}(i\nu_{n})$ because of eq. (\ref{eqn:self_con_for_S}) and eq. (\ref{eqn:dfdh}).
Finally, we can solve for $\chi_{or}(i\nu_{n})$ by combining eq.(\ref{eqn:self_con_for_S}) through eq.(\ref{eqn:pfpm})
\begin{eqnarray}
1/\chi_{or}(i\nu_n)
=\chi_{0z}^{-1}({\bf Q}_{\mathrm{or}},i\nu_n)
+\delta J^{z}_{{\bf Q}_{\mathrm{or}}}
+1/\chi_{\mathrm{loc,or} }(i\nu_n).\,\,\,
\label{eqn:chispin}
\end{eqnarray}
We  immediately recognize this result as the C-EDMFT self-consistent equation for the spin susceptibility at the ordering wave vector ${\bf Q}_{\mathrm{or}}$ and spin component z. Most importantly, this result indicates that a non-polarized Weiss field is consistent with the C-EDMFT treatment. We will show in the appendix how the additional particle hole bubble contribution to the expression for $\chi_{or}$, forbidden by the absence of special $\tilde{{\bf q}}$'s, is introduced if we allow the Weiss field to be polarized. These considerations have important implications for 
the incorporation of superconductivity 
into the formalism,
which we turn to
in the following section.

\section{Superconductivity}
\label{sec:Sup}
We now apply the approach
to the study of superconductivity, 
with the pairing driven
 by the spin-spin interaction in equation (\ref{eqn:ham}).  
 We first focus on an 
 AF
 Ising interaction with $J^z_{ij}\equiv J_{ij}>0$ and set the other $J_{ij}^{\alpha}$ to zero.  
 The  AF interaction $J_{ij}$ 
 favors pairing of electrons with opposite spins;
 a parallel consideration can be made for the case of ferromagnetic interactions, which favors pairing 
 between electrons of the same spins.
 We then discuss the case of a full 
 AF
 Heisenberg interaction, where $J^{a}_{ij}\equiv I_{ij}>0$
are the same
 for $a=1,2,3$.  

Following the discussion of magnetic order in section \ref{sec:mag-order}, 
in the context of pairing, the absence of special $\tilde{{\bf q}}$'s implies we should not include
a contribution from the particle particle bubble in the susceptibility.
This amounts to not allowing the conduction electron band to become ``polarized''
by a finite superconducting order parameter.  
In the appendix, we also discuss
what happens when the conduction electrons are allowed to be pair-polarized.

\subsection{Ising Spin Interaction}
\label{subsec:ising}
Conceptually, we would like to keep the strong inter site interactions (that give rise to the dynamic bosonic bath) while promoting a single mode in the static pairing channel in order to give it the chance to condense.  We do so, by singling out the static, attractive pairing interaction between the paring operators $\hat {\Delta}^{\dagger}_{i\sigma j\bar{\sigma}}$ and $\hat {\Delta}_{i\sigma j\bar{\sigma}}$ [defined as $\hat {\Delta}^{\dagger}_{i\sigma j\bar{\sigma}}(\tau)=c_{i\sigma}^{\dagger}(\tau) c_{j\bar{\sigma}}^{\dagger}(\tau)$].  We rewrite the spin-spin interaction as
\begin{eqnarray}
&&\int_{0}^{\beta}d\tau \sum_{\langle i,j \rangle} J_{ij} S_{i}^{z}(\tau) S_{j}^{z} (\tau)
\nonumber
\\ 
&=&-\sum_{\langle i,j\rangle,\sigma}\frac{J_{ij}}{4\beta} \sum_{\omega} \hat{\Delta}^{\dagger}_{i\sigma j\bar{\sigma}}(i\omega)\hat{\Delta}_{i\sigma j\bar{\sigma}}(i\omega)
\nonumber
\\ 
&+&\sum_{\langle i,j\rangle} \frac{1}{\beta^3}\sum_{\omega,\omega_1,\omega_2} \frac{J_{ij}}{4}(1-\delta_{\alpha,\bar{\gamma}})c^{\dagger}_{i\alpha}(i\omega_{1}-i\omega) 
\nonumber
\\ 
&\times & \sigma_{\alpha\beta}^{z}c_{i\beta}(i\omega_{1})c^{\dagger}_{j\gamma}(i\omega_{2}+i\omega)\sigma_{\gamma\delta}^{z}c_{j\delta}(i\omega_{2})\label{eqn:pair_sep}
\\
&=&-\sum_{\langle i,j\rangle,\sigma}\frac{J_{ij}}{4\beta} \hat{\Delta}^{\dagger}_{i\sigma j\bar{\sigma}}(i\omega=0)\hat{\Delta}_{i\sigma j\bar{\sigma}}(i\omega=0)
\nonumber
\\ 
&+&\sum_{\langle i,j\rangle} \frac{1}{\beta^3}\sum_{\omega,\omega_1,\omega_2} \frac{J_{ij}}{4}(1-\delta_{\omega_{1},-\omega_{2}}\delta_{\alpha,\bar{\gamma}})c^{\dagger}_{i\alpha}(i\omega_{1}-i\omega) 
\nonumber
\\ 
&\times & \sigma_{\alpha\beta}^{z}c_{i\beta}(i\omega_{1})c^{\dagger}_{j\gamma}(i\omega_{2}+i\omega)\sigma_{\gamma\delta}^{z}c_{j\delta}(i\omega_{2}),
\label{eqn:freq_sep}
\end{eqnarray}
where the repeated indices $\alpha,\beta,\gamma,\delta$ are summed over. Here in the first step we separated out the pairing interaction, and then further separated out the static mode of the pairing interaction in the second step. We then introduce a Hubbard-Stratonovich transformation to decouple the attractive interaction in the pairing channel.
\begin{eqnarray}
&\exp&\left(\sum_{\langle i,j\rangle,\sigma}\frac{J_{ij}}{4\beta} \hat{\Delta}^{\dagger}_{i\sigma j\bar{\sigma}}(i\omega=0)\hat{\Delta}_{i\sigma j\bar{\sigma}}(i\omega=0) \right)
\nonumber
\\
&=&\int \mathcal{D}[\Delta,\bar{\Delta}]\exp\Bigg(-\sum_{\langle i,j \rangle,\sigma} \Big[ \-\frac{ 4\beta|\Delta_{i\sigma,j\bar{\sigma}}|^{2}}{J_{ij}}
\nonumber
\\
&+&\int_{0}^{\beta}d\tau \Delta_{i\sigma,j\bar{\sigma}}\hat{\Delta}^{\dagger}_{i\sigma j\bar{\sigma}}(\tau)
+\bar{\Delta}_{i\sigma,j\bar{\sigma}}\hat{\Delta}_{i\sigma j\bar{\sigma}}(\tau) \Big] \Bigg).
\label{eqn:HS}
\end{eqnarray}
We note that if we choose to decouple the first term of equation (\ref{eqn:pair_sep}) -- instead of that of equation (\ref{eqn:freq_sep}) -- and then make the approximation that the Hubbard-Stratonovich field is constant as a function of $\tau$ (i.e. keeping only the static mode of the 
Hubbard-Stratonovich field), we would arrive at exactly the RHS of equation (\ref{eqn:HS}). Thus we see that by separating out the static mode before rather than after the Hubbard-Stratonovich decoupling, the non-static modes of the pairing interaction will be absorbed in the second term in equation (\ref{eqn:freq_sep}). This term is interpreted as the remaining spin-spin interaction. 

Now we take the saddle point approximation of $\Delta$ and follow the steps in section~\ref{subsec:ECTR} to carry out a generalized cavity construction.  Up to 
additive
constants, we obtain the effective cluster action, 
\begin{eqnarray}
& S_{C,I}&=S_{C,I}^{0} \hspace{8mm}
\nonumber
\\
&-&\sum_{\langle X,Y \rangle,\sigma}\int_0^{\beta} d\tau\left(\Delta_{X\sigma Y \bar{\sigma}} \hat{\Delta}^{\dag}_{X\sigma Y \bar{\sigma}} (\tau)+ \mathrm{h.c.} \right)
\nonumber
\\
& -& \int_0^{\beta}d\tau d\tau^{\prime} \sum_{X,Y,\sigma} c_{X\sigma}^{\dag}(\tau)\mathcal{G}_{0,XY}^{-1}(\tau-\tau^{\prime})c_{Y\sigma}(\tau^{\prime})\nonumber
\\
&-&\frac{1}{2}\int_{0}^{\beta} d\tau d\tau^{\prime} \sum_{X,Y}S_{X}^z(\tau)\chi^{-1}_{0,XY}(\tau-\tau^{\prime})S_{Y}^z(\tau^{\prime})
\nonumber
\\
&+& \delta S.
\label{eqn:c-SCaction}
\end{eqnarray}
The saddle point equation leads to an additional self consistent equation for the superconducting order parameter
\begin{equation}
\Delta^c_{X_{i}\sigma X_{j}\bar{\sigma}}=\frac{J^c_{X_i X_j}}{4\beta}\int_{0}^{\beta}d{\tau}\langle \hat{\Delta}_{X_{i}\sigma X_{j}\bar{\sigma}}(\tau)\rangle_{C},
\nonumber\\
\label{eqn-gapSC}
\end{equation}
where the average is taken with respect to the effective cluster model in equation (\ref{eqn:c-SCaction}).  
The additional term in the action, $\delta S$ represents all the modifications in the effective action caused 
by separating the zero frequency pairing interaction. The exact expression of $\delta S$ can be found in the Appendix B.
We see from the expression there
that all terms in $\delta S$ are suppressed by factors of
$1/(J\beta)$; therefore, at sufficiently low temperatures 
including the quantum critical regime, $\delta S$ can be safely neglected.
In the following we only consider the low temperature limit and make the approximation that $\delta S \approx 0$.
Working in this approximation, 
the self consistence conditions for both $\mathcal{G}_{0,XY}^{-1}$ and $\chi^{-1}_{0,XY}$ remain the same as in
the previous sections.

It is useful to note that
 this approach can also be formulated in momentum space.  
 Using the DCA Fourier transform defined in equation (\ref{eqn:dcaft}) amounts to replacing $J_{ij}$ 
 by $J_{ij}^{DCA}=N_c/N\sum_{\tbq}e^{i\tbq\cdot(\tilde{{\bf x}}_i-\tilde{{\bf x}}_j)}[J^{\alpha}_{\mathrm{DCA}}
 (\tilde{{\bf q}})]_{{\bf X}_i,{\bf X}_j}$ in the spin channel.  In order to treat the pairing channel on the same footing, 
 we use $J_{ij}^{DCA}$ starting in equation (\ref{eqn:freq_sep}).   This then amounts to replacing $J^c_{X_i X_j}$ 
 in equation (\ref{eqn-gapSC})  with $[J^c_{DCA}]_{X_i X_j}$, which is the coarse grained interaction in cluster 
 momentum $(\bar{J}_{{\bf Q}}$) that is Fourier transformed back to real space cluster variables 
 (see ref.~\onlinecite{cluster-review}) $[J^c_{DCA}]_{X_i X_j}
 =1/N_c\sum_{{\bf Q}}e^{i{\bf Q}\cdot ({\bf X}_i - {\bf X}_j)}\bar{J}_{\bf Q} $.
 In addition, this leads to periodic boundary condition on the cluster in real space.

\subsection{Heisenberg Spin Interaction}
We now consider the case of an
AF
Heisenberg spin spin interaction $I_{ij}$.
The derivation proceeds in parallel with the previous Ising case. 
Here, we separate out the singlet term with an attractive interaction~\cite{baskaran1987}  
\begin{eqnarray}
& &\int_{0}^{\beta}d\tau \sum_{\langle i,j \rangle} I_{ij} \vec{S}_{i}(\tau) \cdot \vec{S}_{j} (\tau)
\nonumber
\\ 
&=&-\sum_{\langle i,j\rangle}\frac{I_{ij}}{\beta}\int_{0}^{\beta} d\tau_{1}d\tau_{2}\hat{\Delta}_{ij}^{\dag}(\tau_1)\hat{\Delta}_{ij}(\tau_2)
\nonumber
\\
&+&\sum_{\underset{\omega,\omega_1,\omega_2}{\langle i,j\rangle}} \frac{I_{ij}}{4\beta^3}
(\sigma_{\mu\beta}^{\alpha}\sigma_{\gamma\nu}^{\alpha}-\delta_{\omega_{1},-\omega_{2}}(2\delta_{\mu\delta}\delta_{\beta\nu}-2\delta_{\mu\beta}\delta_{\gamma\nu}))
\nonumber
\\ 
&\times & c^{\dagger}_{i\mu}(i\omega_{1}-i\omega)c_{i\beta}(i\omega_{1})c^{\dagger}_{j\gamma}(i\omega_{2}+i\omega)c_{j\nu}(i\omega_{2}),
\end{eqnarray}
we have defined the singlet creation operator between sites $i$ and $j$ as
\begin{equation}
\hat{\Delta}^{\dag}_{ij}=\frac{1}{\sqrt{2}}\left(c_{i\uparrow}^{\dagger}c_{j\downarrow}^{\dagger}-c_{i\downarrow}^{\dagger}c_{j\uparrow}^{\dagger}\right),
\label{eqn:pair}
\end{equation}
and its hermitian conjugate $\hat{\Delta}_{ij} =\left(c_{j\downarrow}c_{i\uparrow}-c_{j\uparrow}c_{i\downarrow}\right) /\sqrt{2}$.
As before, we then introduce a static Hubbard-Stratonovich field to decouple the singlet interaction $\hat{\Delta}_{ij}^{\dag}\hat{\Delta}_{ij}$,
the saddle point equation for $\Delta_{ij}$ now becomes
\begin{equation}
\Delta_{X_{i}X_{j}}=\frac{I_{c X_iX_j}}{\beta}\int_{0}^{\beta}d{\tau}\langle \hat{\Delta}_{X_{i}X_{j}}(\tau)\rangle_{C}.
\end{equation}
with an effective action,
\begin{eqnarray}
& S_{C,H}&= S_{C,H}^{0} \hspace{63mm}
\nonumber
\\
 &-&\sum_{\langle X,Y \rangle}\int_0^{\beta} d\tau\left(\Delta_{X Y } \hat{\Delta}^{\dag}_{XY } (\tau)+ \mathrm{h.c.} \right)
\nonumber
\\
&-& \int_0^{\beta}d\tau d\tau^{\prime} \sum_{X,Y,\sigma} c_{X\sigma}^{\dag}(\tau)\mathcal{G}_{0,XY}^{-1}(\tau-\tau^{\prime})c_{Y\sigma}(\tau^{\prime})
\nonumber
\\
&-& \frac{1}{2}\int_0^{\beta}d\tau d\tau^{\prime} \sum_{\underset{a\neq0}{X,Y}}S^{a}_X(\tau)\chi_{0a,XY}^{-1}(\tau-\tau^{\prime})S^{a}_Y(\tau^{\prime}).
\nonumber\\
\label{eqn:c-action-general}
\end{eqnarray}
Again, we have not allowed the single particle Weiss field to become polarized from the finite superconducting order parameter. The Heisenberg isolated cluster action $S_{C,H}^0$, is equation (\ref{eqn:isolated-action}) with ${\bf J}_{c}^{\alpha}=0$ for $\alpha=0$ and ${\bf J}_{c}^{\alpha}={\bf I}_{c}$ for $\alpha=1,2,3$. 
We have also ignored the additional part of the action $\delta S_H$ that is suppressed by a factor of $1/(J \beta)$ (see the appendix for a discussion of the Ising case).  
We remark that it is
possible to use this formalism to describe states that have both magnetic order and superconductivity by including a finite $h_{\mathrm{loc}}$ as described in section~\ref{sec:mag-order}.

\subsection{Pairing Susceptibility}
\label{sec:normal-state}
In this section we derive the zero momentum lattice pairing susceptibility $\chi_{SC}(i\nu_{n})\equiv\chi_{\mathrm{pair}}^{\mathrm{lat}}({\bf q}=0,i\nu_{n})$ defined as
\begin{eqnarray}
\chi_{SC}({i\nu_{n}})&= &\frac{1}{N(z/2)}\sum_{\langle i,j\rangle,\sigma}\sum_{\langle k,l\rangle,\lambda}f^*_{i,j}f_{k,l}g^*_{\sigma\bar{\sigma}}g_{\lambda\bar{\lambda}}
\nonumber
\\ 
&\times&\int_0^{\beta}d\tau \langle T_{\tau} :\hat{\Delta}_{i\sigma j\bar{\sigma}}(\tau): :\hat{\Delta}^{\dag}_{k\lambda l\bar{\lambda}}:\rangle e^{i\nu_{n}\tau}, \nonumber\\
\end{eqnarray}
where $N(z/2)$ is the number of bonds in the lattice, with $z$ being the number of nearest neighbors, and $f_{i,j}$, $g_{i,j}$ are the pairing symmetry factors~\onlinecite{Mineev-book} in real and spin space respectively.
We will focus on the case of an Ising spin interaction $J_{ij}$ but this can be easily generalized to the case of 
a Heisenberg interaction.  We first project the superconducting gap onto a particular symmetry channel assuming 
the gap amplitude $\Delta_0$ is uniform across each bond, i.e. 
$\Delta_{X\sigma Y\bar{\sigma}}=f_{XY}g_{\sigma\bar{\sigma}}\Delta_0$ 
(and complex conjugate $\Delta_{X\sigma Y\bar{\sigma}}^*=f_{XY}^*g_{\sigma\bar{\sigma}}^*\Delta_0^*$), 
where the phase factor in real space is given by $f_{XY}$ and that in spin space is $g_{\sigma\bar{\sigma}}$, 
and they have the property 
$|f_{XY}|^2,|g_{\sigma\bar{\sigma}}|^2=1$ (see ref.~\onlinecite{Mineev-book}). 
The procedure is similar to that
described in section~\ref{sec:LT} for magnetism, 
although here we have to project into a particular symmetry channel (through $f_{ij},g_{\sigma\bar{\sigma}}$).
We arrive at the following expression for the zero momentum lattice pairing susceptibility.
\begin{equation}
\chi_{SC}(i\nu_{n})=\frac{1}{1/\chi^{\loc}_{\mathrm{pair}}(i\nu_{n}) -J_{SC}},
\label{eqn:chipair}
\end{equation}
where we have defined the effective pairing interaction $1/J_{SC} = \frac{4}{N_b}\sum_{\langle X_i,X_j \rangle,\sigma}1/J_{X_iX_j}^c$, and the cluster pairing susceptibility $\chi_{\mathrm{pair}}^{\mathrm{loc}}(i\nu_{n})\equiv\chi_{\mathrm{pair}}^{\mathrm{loc}}({\bf Q}=0,i\nu_{n})$, where
\begin{eqnarray}
\chi_{\mathrm{pair}}^{\mathrm{\loc}}(i\nu_{n})&=&\frac{1}{N_b}\sum_{\langle X,X^{\prime}\rangle,\sigma}\sum_{\langle Y,Y^{\prime}\rangle,\lambda}f^*_{X,X^{\prime}}f_{Y,Y^{\prime}}g^*_{\sigma\bar{\sigma}}g_{\lambda\bar{\lambda}}
\nonumber
\\ 
&\times&\int_0^{\beta}d\tau \langle T_{\tau} :\hat{\Delta}_{X\sigma X^{\prime}\bar{\sigma}}(\tau): :\hat{\Delta}^{\dag}_{Y\lambda Y^{\prime}\bar{\lambda}}:\rangle_C e^{i\nu_{n}\tau}
\nonumber\\
\end{eqnarray}
where $N_b = \sum_{\langle X,Y\rangle}$ is the number of individual bonds in the cluster. 
Previous treatments of two particle response functions in various cluster theories demands much more
computational effort because it involves the inversion of the Bethe-Salpeter equation, 
which is in principle a matrix equation of infinite dimension in the space of three wave vectors and frequencies.  
In our approach, the cluster susceptibilities completely determine the corresponding lattice quantities.

We note that this can also be formulated in momentum space after the symmetry factors are first coarse grained in momentum space $\bar{f}({\bf K}) = N_c/N\sum_{\tbk} f(\tbk + {\bf K})$ (where $f({\bf k})$ is the Fourier transform of $f_{ij}$) and then Fourier transformed to real cluster space $\bar{f}_{X_i X_j}=[f_{DCA}]_{X_i X_j}=1/N_c\sum_{{\bf K}}e^{i{\bf K}\cdot ({\bf X}_i - {\bf X}_j)}\bar{f}_{\bf K} $. In this case, the coarse grained symmetry factors no longer have to satisfy $|\bar{f}_{X_i X_j}|^2=1$, and as a result the effective pairing interaction becomes $1/J_{SC} = \frac{4}{N_b}\sum_{\langle X_i,X_j \rangle,\sigma}|\bar{f}_{X_iX_j}|^2/J_{X_iX_j}^c$.

In the appendix, we contrast this expression with the result of allowing the single particle Weiss field to acquire anomalous terms.  In this case, we find additional contributions, corresponding to the particle particle bubble which only contributes if special $\tilde{\bf{q}}$'s exist.
 
\section{Effective Cluster Models}
\label{sec:ECM}
The formalism we have discussed so far also applies
to a variety of strongly correlated problems 
aside
from the Hamiltonian we have been considering in equation (\ref{eqn:ham}).  
One such example is
a two band model, namely
the Anderson lattice Hamiltonian appropriate for the description of heavy fermion materials~\cite{SiSteglich_Science2010}.  
The model describes a band of conduction electrons hybridized with a band of localized, 
highly correlated $f$-electrons and is defined as
\begin{eqnarray}
H_{\mathrm{AL}} &=& \sum_{\langle i,j \rangle, \sigma} t_{ij}(c_{i\sigma}^{\dag}c_{j\sigma} + \mathrm{h.c}) + \sum_{i} \left(\epsilon_f n_{fi} +U n_{fi\uparrow}n_{fi\downarrow}\right)
\nonumber
\\
&+& \sum_{i,\sigma}\left(Vc_{i\sigma}^{\dag}f_{i\sigma} + \mathrm{h.c.} \right) + \sum_{\langle i,j \rangle} J_{ij}S^z_{fi} S^z_{fj} .
\label{eqn:ham-anderson}
\end{eqnarray}
As usual, we have explicitly included an Ising RKKY interaction between the $f$-electron spins.
The RKKY and Kondo interactions compete, and tuning their ratio can lead to a quantum phase transition
between a heavy Fermi liquid and an antiferromagnet.  In certain cases,  
the QCP
is of the SDW type, where the heavy quasiparticles remain intact across the transition
and undergo a SDW transition.
In other cases, the SDW description fails, and the physics of critical Kondo destruction
comes into play.   

Focusing on the normal state properties, applying the extended dynamical cluster theory of section~\ref{subsec:ECTM} 
in momentum space we arrive at the effective cluster action
\begin{eqnarray}
&S_C^{\mathrm{AL}}&=  \int_0^{\beta}d\tau \,\sum_{{\bf Q}} U n_{f{\bf Q}\uparrow}(\tau)n_{f-{\bf Q}\downarrow}(\tau) 
+ \epsilon_f n_{f{\bf Q}}(\tau)   \,\,\,\,\,\,\,\,\,\,\,\,\,
\\
&+&  \int_0^{\beta}d\tau \sum_{{\bf Q}}  \bar{J}_{{\bf Q}}S^z_{f{\bf Q}}(\tau)S^z_{f-{\bf Q}}(\tau)
\nonumber
\\
&-&\int_0^{\beta}d\tau d\tau^{\prime} \sum_{{\bf K},\sigma} f_{{\bf K}\sigma}^{\dag}(\tau)\mathcal{G}_{0,{\bf K}}^{-1}(\tau-\tau^{\prime})f_{{\bf K}\sigma}(\tau^{\prime})
\nonumber
\\
&-&\frac{1}{2}\int_0^{\beta}d\tau d\tau^{\prime} \sum_{{\bf Q}} S^z_{f{\bf Q}}(\tau)\chi_{0,{\bf Q}}^{-1}(\tau-\tau^{\prime})S^z_{f-{\bf Q}}(\tau^{\prime}).
\nonumber 
\label{eqn:c-ALaction}
\end{eqnarray}
For the case of  2-d AF exchange 
fluctuations, the divergence of the spin susceptibility at the ordering wave vector implies [through the self consistent equation (\ref{eqn:self-chi-Q})] 
that the local spin susceptibility with cluster momentum ${\bf Q}_{\mathrm{or}}$ is also (logarithmically) divergent~\cite{Si-2003}.
This leads to a spin Weiss field associated with the critical momentum channel $\chi_0({\bf Q}_{AF},i\nu_n)$ that develops a sub-ohmic spectral density $\mathrm{Im}\chi_0({\bf Q}_{AF},\omega+0^+)\sim \omega^s$.  Based on universality, we can regard the effective cluster model in equation (\ref{eqn:c-ALaction}) with a sub-ohmic density of states for the ordered channel, as an effective model that contains both Kondo destruction and pairing correlations induced from
AF exchange interactions.  For the simplest case of $N_c=2$, and keeping only the critical degrees of freedom [i.e. only the Weiss field in the ordered channel $\chi_0({\bf Q}_{AF},i\nu_n)$] we arrive at a simplified
model to study pairing correlations near a Kondo destroyed QCP.  This model was proposed and solved in ref.~\onlinecite{Pixley-2013} using a combination of continuous time quantum Monte Carlo and the numerical renormalization group.  It was shown that the cluster pairing susceptibility $\chi_{\mathrm{pair}}^{\mathrm{loc}}$ is enhanced at the Kondo destruction QCP.  It will be
important to consider the full self consistent solution and determine if local quantum criticality with Kondo destruction survives finite size cluster corrections and if so, how large $\chi_{SC}$ is near the local QCP.

Another strongly correlated problem of central interest is the extended Hubbard model, which adds to the standard Hubbard model an inter-site density-density interaction.  We note in passing
 that incorporating a spin orbit coupling can lead to topologically non-trivial ground states in the presence of interactions~\cite{Freedman-2005,Witczak-2014}.  The extended Hubbard model is thought to be the appropriate model to describe certain types of organic superconductors, stripe charge order in the cuprates and different types of Mott transitions.  
 In addition, when the
 inter-site interaction is attractive ($V_{ij}<0$),
 the model may also feature
 inter-site pairing and, in the presence of a spin-orbit coupling, topological superconductivity~\cite{Farrel-2014}. 
 It is defined as
\begin{eqnarray}
H&=& \sum_{i,j, \alpha,\beta} t_{ij}^{\alpha\beta}(c_{i\alpha}^{\dag}c_{j\beta} + \mathrm{h.c}) + U\sum_i n_{i\uparrow}n_{i\downarrow}
\nonumber
\\
&+&  \sum_{\langle i,j \rangle} V_{ij}:n_i::n_j:,
\label{eqn:ham4}
\end{eqnarray}
where we have generalized the tight-binding hopping terms to a spin dependent hopping matrix $t_{ij}^{\alpha\beta}$;
this form is sufficiently general to allow
for a spin orbit coupling term and keep 
$i,j$ that are not necessarily
nearest neighbors.  Tuning the ratio of $U/V$ can lead to a variety of quantum phase transitions between 
a Fermi liquid, a band insulator, and a Mott insulator.  
Whereas tuning the strength of the spin orbit coupling can lead to topological transitions between a band and topological insulator.   Applying the formalism of section~\ref{subsec:ECTR} leads to the effective cluster action 
\begin{eqnarray}
&S_C^{\mathrm{EH}}&=\int_0^{\beta}d\tau \sum_{X\in C}Un_{X\uparrow}(\tau)n_{X\downarrow}(\tau)
\nonumber
\\
&+& \int_0^{\beta}d\tau\sum_{\langle X,Y \rangle}V_c :n_{X}(\tau)::n_{Y}(\tau):
\nonumber
\\
&-& \int_0^{\beta}d\tau d\tau^{\prime} \sum_{X,Y,\mu,\nu} c_{X\mu}^{\dag}(\tau)\mathcal{G}_{0,XY\mu\nu}^{-1}(\tau-\tau^{\prime})c_{Y\nu}(\tau^{\prime})
\nonumber
\\
&-&\frac{1}{2}\int_0^{\beta}d\tau d\tau^{\prime} \sum_{X,Y} :n_X(\tau):\chi_{0\mathrm{ch},XY}^{-1}(\tau-\tau^{\prime}):n_Y(\tau^{\prime}): 
\nonumber\\
\label{eqn:c-action-EH}
\end{eqnarray}
Studying the different types of quantum phase transitions in the effective cluster model alone 
[i.e., without implementing the self-consistency]
can lead to significant new insights regarding behavior of the QCPs in the lattice problem.
In other words, the effective cluster model of equation (\ref{eqn:c-action-EH}) serves as a 
simplified 
model that can provide insights into
different types of Mott transitions, superconductivity and even
interacting topological phase transitions.

\section{C-EDMFT solution methods}
\label{sec:Sol_methods}
The self consistent cluster dynamical mean field equations form a set of highly non-linear equations.  
Their solution requires an accurate and reliable solution of the cluster impurity model, which is iteratively solved.
There are analytical tools and computational methods 
that are suitable
for studying the cluster impurity models.
From a computational perspective, 
solving the cluster impurity models 
represents the most extensive efforts of solving the self consistent equations. In the presence of a phase transition, the number of iterations necessary to solve the equations can become quite large due to a ``critical slowing down''.  In this case, it is very useful to use mixing techniques that are well known in the context of density functional theory, in order to reduce the number of iterations that are needed to achieve self consistency~\cite{Zitko-2009}.
 
The cluster model can be solved using,
for example
the exact diagonalization (ED), 
the numerical renormalization group (NRG), the density matrix renormalization group (DMRG) 
and quantum Monte Carlo methods.  

Including bosonic baths in diagonalization based techniques can be done, but it requires a truncation of the infinite bosonic Hilbert space.  In addition, the presence of numerous bosonic baths can make such an approach quite computationally demanding.  However, this does not rule out ED and NRG techniques, provided there is some physical intuition of which bosonic bath is going to drive the system through a quantum phase transition. Then retaining this single bosonic bath is necessary to capture the critical universal properties, while the other baths serve to renormalize the effective model parameters.

The recently developed continuous time quantum Monte Carlo (CT-QMC) has the advantage that the bosons are traced out and the algorithm is numerically exact.  These methods have been adapted to treat scalar bosonic baths that interact with the impurities charge~\cite{Werner-2007,Werner-2010} or spin~\cite{Pixley-2010,Pixley2-2013} degrees of freedom.  
The case of
a vector bosonic bath~\cite{Otsuki-2013} 
has also been studied.

Recently, the CT-QMC has been generalized to a two impurity~\cite{Pixley-2013} model in the presence of a single bosonic bath.  As the additional bosonic baths that arise in the C-EDMFT treatment commute with each other, the method in Ref.~\onlinecite{Pixley-2013} can naturally be used to solve the C-EDMFT equations to high accuracy. Generalizations to larger cluster is still possible within such a framework, provided the algorithm doesn't suffer from a sign problem arising from the fermionic degrees of freedom.
As described in section V, for the minimal case of the two-impurity C-EDMFT approach to the periodic Anderson model,
in the vicinity of an antiferromagnetic QCP,
it would be adequate to keep only a single bosonic bath that is coupled to the staggered combination of the
local-moment spins in the cluster. The method of Ref.~\onlinecite{Pixley-2013} then makes it feasible to study 
the quantum critical behavior of the periodic Anderson model.

\section{Discussion}
\label{sec:Dis}
In this work we have derived a cluster EDMFT formulation by a locator expansion about a dressed 
cluster limit.  
An alternative derivation can also be done using a Baym-Kadanoff functional~\cite{cluster-review,RVB-DMFT}.  
In this case the generating functional of the grand potential is (focusing on one of the two particle channels)
\begin{eqnarray}
\Gamma_{\mathrm{BK}}[{\bf G},\bm{\chi}] &=& \mathrm{Tr}\left[\log ({\bf G})\right]- \mathrm{Tr}\left[{\bf G}(\bf{G}_0^{-1}-{\bf G}^{-1})\right]
\nonumber
\\
&-&\frac{1}{2}\mathrm{Tr}\left[\log (\bm{\chi})\right]+\mathrm{Tr}\left[\bm{\chi}(\bm{\chi}_0^{-1}-\bm{\chi}^{-1})\right]
\nonumber
\\
&+&\Phi[{\bf G},\bm{\chi}],
\end{eqnarray}
with the stationary conditions $\delta \Gamma/\delta {\bf G} = \delta \Gamma/\delta \bm{\chi}=0$.  The self energies are then given by $\delta \Phi[{\bf G},\bm{\chi}]/\delta {\bf G}=\bm{\Sigma}$, $\delta \Phi[{\bf G},\bm{\chi}]/\delta \bm{\chi}={\bf M}$.  Within the cluster approximation, the self energies are calculated from an effective cluster model.  Therefore, the approximation that the functional $\Phi$ is only a functional of ${\bf G}_{\mathrm{loc}}$ and $\bm{\chi}_{\mathrm{loc}}$ leads to the cluster EDMFT equations.  
Therefore,
analogous to the EDMFT approach~\cite{Si-1996,Smith},
the C-EDMFT approach is also
conserving~\cite{Baym-1961}. 

As we have discussed in the introduction, a main success of EDMFT compared to DMFT has been
in its solution to the Kondo lattice Hamiltonian and the theory of local quantum criticality
~\cite{Si_etal_JPSJ2013,Si-2001,Si-2003,Zhu.2003,Glossop.2007,Zhu.2007,Si-2005}.
This rests on treating the RKKY interactions between the local moments in a dynamical fashion,
through the self-consistent bosonic bath. The self-consistent solution yields an unconventional
 QCP with a dynamical
spin susceptibility satisfying $E/T$ scaling and anomalous dynamical critical exponent, similar to 
what has been observed in quantum-critical heavy-fermion 
metals~\cite{Aronson.95+Schroeder.00,Paschen.00+Gegenwart.07+Friedemann.10, Shishido, SiPashchen13}. For this result, it is important that the mapped Bose-Fermi Kondo model itself contains 
a Kondo-destruction QCP ~\cite{Zhu-02,Pixley2-2013}.
Studies of the two impurity Bose-Fermi Anderson model in Ref.~\onlinecite{Pixley-2013}
have shown that a Kondo-destruction QCP persists, with similar scaling properties for the
staggered dynamical spin susceptibility. This raises the prospect that the C-EDMFT approach
discussed here will be able to study the interplay between the unconventional quantum critical normal 
state and superconductivity. As discussed earlier in the paper, such an interplay is important to the 
understanding of heavy fermion superconductivity. 
We note that the dynamical treatment of the RKKY interactions differentiates EDMFT and C-EDMFT 
from DMFT, which does not treat the RKKY interactions beyond a static mean field approximation,
as well as the cluster-DMFT~\cite{Leo-2008}, which treats the RKKY interactions
within the cluster in a dynamical fashion, but does not do so for those interactions outside of a chosen cluster.

\section{Conclusions}
\label{sec:Con}
In this manuscript we have presented a new cluster extended dynamical mean field approach.  We have developed the equations in both real and momentum space, incorporating magnetic order and superconductivity.  We have also determined the form of the superconducting correlation functions in the normal state.
We have then used the formalism to derive effective cluster models that are relevant to heavy fermion metals and Mott-Hubbard systems. 
In particular, this formulation for
unconventional superconductivity 
is expected to play a central role to the study of 
quantum critical heavy fermion metals.

{\it Acknowledgements.~} 
We would like to thank Kevin Ingersent and Lili Deng for useful discussions.
This work has been supported by the NSF Grant No. DMR-1309531, the Robert A.\ Welch Foundation
Grant No.\ C-1411 (J.H.P., A.C. \& Q.S.), the Alexander von Humboldt Foundation (Q.S.), and the East-DeMarco fellowship (J.H.P.).
One of us (Q. S.)
acknowledges the hospitality of the Aspen Center for Physics (NSF Grant No. 1066293),
the Institute of Physics of Chinese Academy of Sciences,
and the Karlsruhe Institute of Technology.

\section{Appendix A}
\label{sec:appendix-A}
In this appendix we derive the expression for $\delta S$, defined in equation (\ref{eqn:c-SCaction}).  We use the generalized cavity approach to derive the effective action $S_C$,
\begin{equation}
S_{C}=S_{{\bf o}}-\sum_{n=0}^{\infty}\frac{1}{n!}\langle (-\Delta S)^{n} \rangle^{({\bf o})}_{\mathrm{connected}}
\end{equation}
We have defined $\Delta S$ as the part of the action that connects cluster ${\bf o}$ to its neighbors.  
The expectation value $\langle \dots \rangle^{({\bf o})}$ is taken with respect to the action $S^{({\bf o})}$, 
which is defined as the action with cluster ${\bf o}$ removed, and $S_{\bf o}$ corresponds to the action of the isolated cluster ${\bf o}$. Similar to DMFT, after rescaling $\delta t$ and $\delta J$, in the limit of large coordination, only the $n=1$ and $n=2$ terms contribute. 
In the remainder of this appendix, 
 we omit the ``connected'' label and take cluster ${\bf o}$ to be equal to ${\bf r}_0={\bf o}$.

We separate $\Delta S$ into three pieces
\begin{align}
\Delta{S}&=\int_{0}^{\beta}d\tau\sum_{{\bf r}_i,XY}\delta J({\bf r}_{0}-{\bf r}_{i})_{XY}S^{z}_{{\bf r}_{0} X}(\tau)S^{z}_{{\bf r}_{i}Y}(\tau)
\nonumber
\\
&+\int_{0}^{\beta}d\tau_{1}d\tau_{2}\sum_{{\bf r}_i,X,Y,\sigma}\frac{\delta J({\bf r}_{0}- {\bf r}_{i})_{XY}}{4\beta} 
\nonumber
\\
&\times c_{{\bf r}_{0}X\sigma}^{\dagger}(\tau_{1})c_{{\bf r}_{0}X\sigma}(\tau_{2})c_{{\bf r}_{i}Y\bar{\sigma}}^{\dagger}(\tau_{1})c_{{\bf r}_{i}Y\bar{\sigma}}(\tau_{2})
\nonumber
\\
&+\int_{0}^{\beta}d\tau\sum_{i \sigma}c_{{\bf r}_{0}X \sigma}^{\dagger}(\tau) \delta {t} ({\bf r}_{0}-{\bf r}_{i}) c_{{\bf r}_{i}Y\sigma}(\tau)
\nonumber
\\
&\equiv \Delta S_{1}+\Delta S_{2}+\Delta S_{3},
\end{align}

Since there is no interference between the one particle sector and two particle sector in the expansion, we can ignore $\Delta S_{3}$ in the calculation of $\delta S$. We first note that $\langle \Delta S_{1} \rangle^{({\bf o})} $ vanishes, since we are considering a case with no magnetic order. The expectation value $ \langle\Delta (S_{1})^{2} \rangle^{({\bf o})}$ gives the standard expression for the spin Weiss field (see equation (\ref{eqn:cav-chi}) of the main text). The rest of the terms combined give $\delta S$.
\begin{equation}
\delta S= \langle \Delta S_{2} \rangle^{({\bf o})}-\langle \Delta S_{1}\Delta S_{2} \rangle^{({\bf o})}-\frac{1}{2}\langle (\Delta S_{2})^{2} \rangle^{({\bf o})}
\end{equation}
The expression for each term is listed below,
\begin{widetext}
\begin{equation}
\langle \Delta S_{2} \rangle^{({\bf o})}=-\sum_{{\bf r}_i,X,Y,\sigma}\frac{\delta J( {\bf r}_{0}-{\bf r}_{i})_{XY}}{4\beta}\int_{0}^{\beta} d\tau_{1} d\tau_{2} c^{\dagger}_{{\bf r}_{0} X\sigma}(\tau_{1})\langle c_{{\bf r}_{i}Y\bar{\sigma}}(\tau_{2})c^{\dagger}_{{\bf r}_{i}Y\bar{\sigma}}(\tau_{1})\rangle ^{({\bf o})}c_{{\bf r}_{0}X\sigma}(\tau_{2}),
\end{equation}

\begin{align}
\langle T_{\tau} \Delta S_{1}\Delta S_{2} \rangle^{({\bf o})}=\sum_{{\bf r}_i,{\bf r}_j}\sum_{X,Y,Z,W}\sum_{\sigma,\alpha}\frac{\delta J({\bf r}_{0}-{\bf r}_{i})_{XY}\delta J({\bf r}_{0}-{\bf r}_{j})_{ZW}}{4\beta}\int_{0}^{\beta} d\tau_{1} d\tau_{2} d\tau_{3}   S_{{\bf r}_{0}X}^{z}(\tau_{1})c^{\dagger}_{{\bf r}_{0}Z\sigma}(\tau_{2})c_{{\bf r}_{0}Z\sigma}(\tau_{3})
\nonumber
\\
\times \langle T_{\tau} c_{{\bf r}_{i}Y\alpha}^{\dagger}(\tau_{1})c_{{\bf r}_{i}Y\alpha}(\tau_{1})c_{{\bf r}_{j}W\bar{\sigma}}^{\dagger}(\tau_{2})c_{{\bf r}_{j}W\bar{\sigma}}(\tau_{3})\rangle ^{({\bf o})}\sigma_{\alpha\alpha}^{z}
\end{align}

\begin{align}
\langle (\Delta S_{2})^{2} \rangle^{({\bf o})}&=\sum_{{\bf r}_i,{\bf r}_j}\sum_{X,Y,Z,W}\sum_{\alpha,\beta}\frac{\delta J({\bf r}_{0}-{\bf r}_{i})_{XY}\delta J({\bf r}_{0}-{\bf r}_{j})_{ZW}}{16\beta^{2}}\int_{0}^{\beta} d\tau_{1} d\tau_{2} d\tau_{3} d\tau_{4}c^{\dagger}_{{\bf r}_{0}X\alpha}(\tau_{1})c_{{\bf r}_{0}X\alpha}(\tau_{2})c^{\dagger}_{{\bf r}_{0}Z\beta}(\tau_{3})c_{{\bf r}_{0}Z\beta}(\tau_{4})\nonumber
\\
&\times\langle  T_{\tau} c^{\dagger}_{{\bf r}_{i}Y\bar{\alpha}}(\tau_{1})c_{{\bf r}_{i}Y\bar{\alpha}}(\tau_{2}) c^{\dagger}_{{\bf r}_{j}W\bar{\beta}}(\tau_{3})c_{{\bf r}_{j}W\bar{\beta}}(\tau_{4})\rangle^{({\bf o})}
\end{align}
\end{widetext}
The effect of $ \langle \Delta S_{2} \rangle^{({\bf o})}$ is to modify the one particle Weiss field. The other two terms modify the interactions by generating a general two particle interaction which is nonlocal in time. In contrast to the standard Weiss fields [as in equation (\ref{eqn:c-action})], each term contains an additional factor of $1/\beta$, and the last term carries a factor of $1/\beta^2$.  This implies each term is suppressed by at least a factor of $1/(J\beta)$ relative to the standard Weiss fields, and $1/(J\beta)$ can serve as a small parameter for sufficiently low temperatures. In the zero temperature limit, all three terms vanish and do not affect any of the quantum critical properties.

\section{Appendix B}
Here we discuss when happens when the single particle Weiss fields become polarized 
by a finite magnetic or superconducting order parameter.  In the following, 
we focus on the momentum space formulation. We show that the 
additional terms (particle hole or particle particle bubble contributions) come from a polarized single particle Weiss field.
Our considerations here parallel those for EDMFT discussed in ref.~\onlinecite{Si-2005}.

\subsection{Magnetism}
We first discuss the case of magnetic order, with an order parameter finite in the $z$-direction.  Allowing the magnetic order parameter to polarize the single particle Weiss field we obtain an effective cluster action
\begin{eqnarray}
&S_C&= S_C^0 - \int_0^{\beta}d\tau h_{\mathrm{loc}} S^z_{{\bf Q}_{\mathrm{or}}}(\tau) \,\,\,\,\,\,\,\,\,\,\,\,\,\,\,\,\,\,\,\,\,\,\,\,\,\,\,\,\,\,\,\,\,\,\,\,\,\,\,\,\,\,\, \,\,\,\,\,\,\,\,\,\,\,\,\,\,\,\,\,
\nonumber
\\
&-& \int_0^{\beta}d\tau d\tau^{\prime} \sum_{{\bf K},\sigma} c_{{\bf K}\sigma}^{\dag}(\tau)\mathcal{G}_{0,{\bf K}\sigma}^{-1}(\tau-\tau^{\prime})c_{{\bf K}\sigma}(\tau^{\prime})
\nonumber
\\
&-&\frac{1}{2}\int_0^{\beta}d\tau d\tau^{\prime} \sum_{{\bf Q},\alpha} S^{\alpha}_{{\bf Q}}(\tau)\chi_{0\alpha,{\bf Q}}^{-1}(\tau-\tau^{\prime})S^{\alpha}_{-{\bf Q}}(\tau^{\prime}),
\label{eqn:c-action3}
\end{eqnarray}
where now the single particle Weiss field $\mathcal{G}_{0,{\bf K}\sigma}^{-1}$ is different for different spin components and $h_{\mathrm{loc}}$ is given by equation (\ref{eqn:hloc-Q}).
As a result, the expression for the lattice spin susceptibility has changed, focusing on the static spin susceptibility we obtain
\begin{equation}
\tilde{\chi}({\bf Q}_{\mathrm{or}},i\nu_{n}=0)=\frac{1 + \delta I_h }{1/\chi_{or}(i\nu_{n}=0)-\delta I_M}
\label{eqn:rord}
\end{equation}
where $\delta I_M$ is  
\begin{eqnarray}
&\delta I_M& =  \int_0^{\beta} d\tau d\tau^{\prime} \sum_{{\bf K},\sigma} \langle T_{\tau} c_{{\bf K}\sigma}^{\dag}(\tau) c_{{\bf K}\sigma}(\tau^{\prime}) S^z_{{\bf Q}_{\mathrm{or}}}\rangle_C
\nonumber
\\
&\times & 
 \frac{\partial }{\partial M}   \mathcal{G}_{0,{\bf K}\sigma}^{-1}(\tau-\tau^{\prime})[\chi_{\loc}(i\nu_{n}=0)]^{-1}.
\end{eqnarray}
In addition, we have defined $\delta I_h$, which is equivalent to $\delta I_M$, with $M$ replaced with $\sqrt{N_{c}/N}h_{{\bf Q}_{\mathrm{or}},n=0}$ and the derivative is evaluated at $h_{{\bf Q}_{\mathrm{or}}}=0$.

We can compare our result to that of DCA~\cite{cluster-review}, and conclude that the additional contribution $\delta I$ 
comes from the particle hole contribution associated with the special $\tilde{{\bf q}}$'s.

\subsection{Superconductivity}
We now consider allowing the finite superconducting order parameter to ``polarize'' the single particle Weiss fields, i.e. introduce anomalous terms.
We take the saddle point approximation of $\Delta$ and carry out a generalized cavity construction .  Up to additional constants, we obtain the effective cluster action
\begin{eqnarray}
S_{C,I}&=&S_{C,I}^{0}-\int_{0}^{\beta} d\tau d\tau^{\prime}\Psi^{\dagger}(\tau) \bm{\mathcal{G}}^{-1}_{0\Psi}(\tau-\tau^{\prime})\Psi(\tau^{\prime})
\nonumber
\\
&-&\frac{1}{2}\int_{0}^{\beta} d\tau d\tau^{\prime} \sum_{X,Y}S_{X}^z(\tau)\chi^{-1}_{0,XY}(\tau-\tau^{\prime})S_{Y}^z(\tau^{\prime})
\nonumber
\\
&+& \delta S
\label{eqn:c-SCaction2}
\end{eqnarray}
where the Ising isolated cluster action $S_{C,I}^{0}$ is defined in equation (\ref{eqn:isolated-action}) with ${\bf J}_{c}^{\alpha}=0$ for $\alpha\neq z$.  We have defined the Nambu spinor 
$
\Psi^{\dagger}=(c_{X_{1}\uparrow}^{\dagger},...c_{X_{N_{c}}\uparrow}^{\dagger},
c_{X_{1}\downarrow},...,c_{X_{N_{c}}\downarrow}),
$
 and we adopt the Nambu-Gorkov formalism: $G_{\Psi{\mathrm{loc}}}(\tau)=-\langle T_{\tau} \Psi(\tau) \Psi^{\dagger}\rangle$. We use the subscript $\Psi$ to label a $2\mathrm{x}2$ matrix in Nambu space where each element is a matrix in cluster indices.  Now the single particle Weiss field has additional anomalous terms
\begin{eqnarray}
\bm{\mathcal{G}}_{0\Psi}^{-1}(i\omega_n) = 
\left(\begin{array}{c c}
\bm{\mathcal{G}}_0^{-1}(i\omega_n)&  \bm{\mathcal{F}}_0^{-1}(i\omega_n)\\
 \bm{\mathcal{F}}_0^{*-1}(i\omega_n)&  -{\bm{\mathcal{G}}_0^{-1}}^{T}(-i\omega_n)
\end{array}\right).
\end{eqnarray}
The spin Weiss field $\chi^{-1}_{0,XY}(\tau-\tau^{\prime})$ assumes the same form as equation (\ref{eqn:cav-chi}) 
when expressed in terms of the cavity correlation function. As we have discussed in the main text, the additional term 
in the action, $\delta S$ represents all the modifications in the effective action caused by separating 
the zero frequency pairing interaction. Again, in the following, we only consider the low temperature limit and 
make the approximation that $\delta S \approx 0$.

To determine the one particle Weiss field $\bm{\mathcal{G}}_{0\Psi}^{-1}$, we perform a cumulant expansion in inter cluster interactions.  Similar to the inter-site interactions, we begin by separating $\Delta_{i\sigma,j\bar{\sigma}}$ also into intra and inter cluster parts $\bm{\Delta}_{\sigma\bar{\sigma}}({\bf r}_i - {\bf r}_j) =  \bm{\Delta}_{c\sigma\bar{\sigma}} \delta_{{\bf r}_i ,{\bf r}_j} + \delta \bm{\Delta}_{\sigma\bar{\sigma}}({\bf r}_i - {\bf r}_j)$.  This naturally arises from a locator expansion in ${\bf \delta J}$ after we rescale $\delta \bm{\Delta}_{\sigma\bar{\sigma}}({\bf r}_i - {\bf r}_j)$ by ${\bf \delta J}({\bf r}_i - {\bf r}_j)$, which leads to a single particle Greens function in Nambu space 
\begin{eqnarray}
{\bf G}_{\Psi}(\tbk,i \omega_n) &=& \left[{\bf C}_{G_{\Psi}}^{-1}(i \omega_n) -\delta{\bf T}_{\Psi}(\tbk) \right]^{-1}.
\end{eqnarray}
The single particle cumulant is now
\begin{equation}
{\bf C}_{G_{\Psi}}^{-1}(i \omega_n) = (i\omega_n + \mu)\bm{\tau_{3}}\otimes\bm{1} - {\bf T}_{c\Psi} - \bm{\Sigma}_{\Psi}(i\omega_n)
\end{equation}
where $\tau_{3}$ is the $z$-Pauli matrix in Nambu space, together with the following definition for the generalized intra cluster hopping matrix
\begin{equation}
{\bf T}_{c\Psi}=
\left(\begin{array}{c c}
{\bf t}_{c}&  -{\bf{\Delta}}_{c\uparrow\downarrow}\\
 -\bar{\bf{\Delta}}_{c\downarrow\uparrow}&  -{\bf t}^{T}_{c}
\end{array}\right),
\end{equation}
and the inter cluster hopping matrix 
\begin{equation}
\delta {\bf T}_{\Psi}(\tbk)=
\left(\begin{array}{c c}
\delta {\bf t}(\tbk) &  -\delta{\bm{\Delta}}_{\uparrow\downarrow}(\tbk)\\
 -\delta\bar{\bm{\Delta}}_{\downarrow\uparrow}(\tbk) &  -{\delta {\bf t}}^{T}(\tbk)
\end{array}\right).
\end{equation}
Now the generalized one particle Weiss field takes the form
\begin{eqnarray}
\bm{\mathcal{G}}_{0\Psi}^{-1}(i\omega_n) &=& \bm{\Sigma}_{\Psi}(i\omega_n) + {\bf G}^{-1}_{\Psi\mathrm{loc}}(i\omega_n)
\end{eqnarray}
and the self consistency condition becomes
\begin{equation}
{\bf G}_{\Psi\mathrm{loc}}(i\omega_n) = \frac{N_c}{N}\sum_{\tbk}{\bf G}_{\Psi}(\tbk,i\omega_n).
\end{equation}
The superconducting order parameter is then determined self consistently from the saddle point value:
\begin{equation}
\Delta_{cX_{i}\sigma X_{j}\bar{\sigma}}=\frac{J_{c,X_i X_j}}{4\beta}\int_{0}^{\beta}d{\tau}\langle \hat{\Delta}_{cX_{i}\sigma X_{j}\bar{\sigma}}(\tau)\rangle_{C}
\end{equation}

In principle, for a real space cluster scheme such as CDMFT, the translation invariance inside the cluster is broken since the couplings on the boundary are now treated different than those inside the cluster. Thus, the order parameter $\Delta_{cX_{i}\sigma X_{j}\bar{\sigma}}$ may in principle take different value on different bonds. We could obtain an estimate of the pairing amplitude $\Delta_{0}$ by averaging $\Delta_{cX_{i}\sigma X_{j}\bar{\sigma}}$ over each bond (note that the $ 2\times2$ is a special case where the four sites are indeed equivalent, and such a procedure is unnecessary). After that $\delta \bm{\Delta}_{\sigma\bar{\sigma}}({\bf r}_i - {\bf r}_j)$ is then constructed using the translation invariance of the lattice. 
Two major differences between these self consistent equations and those in CDMFT~\cite{Haule-2007} are the explicit appearance of the order parameter and the fact that the inter-site magnetic interaction is driving the superconducting pairing.

\subsection{Pairing Susceptibility}
Considering the case of an Ising interaction and following section~\ref{sec:normal-state}, we focus on the momentum construction, but we write all of the terms in real cluster space to make the connection to our previous discussions more explicit.
Focusing on the static pairing susceptibility, allowing the single particle Weiss field to acquire anomalous terms, we find an additional contribution to the paring susceptibility,
\begin{equation} 
\tilde{\chi}_{\mathrm{pair}}({\bf q=0},i\nu_{n}=0) 
=\frac{1+\delta \Gamma_{h_p}}{1/\chi_{SC}(i\nu_{n}=0)-J_{SC}\delta \Gamma_{\Delta_0}} .
\end{equation}
Here, the additional contribution is 
\begin{eqnarray}
&\delta \Gamma_{\Delta_0}& = \frac{1}{\beta}\int_0^{\beta}d\tau_1d\tau_2d\tau_3 \sum_{\langle X, Y \rangle,\sigma}\sum_{X^{\prime},Y^{\prime}}
\nonumber
\\ 
&\times&
\bar{f}^*_{XY}g^*_{\sigma\bar{\sigma}} \langle T_{\tau} c_{X\bar{\sigma}}(\tau_1)c_{Y\sigma}(\tau_1) c_{X^{\prime}\uparrow}^{\dag}(\tau_2)c_{Y^{\prime}\downarrow}^{\dag}(\tau_3) \rangle_{C}
\nonumber
\\ 
&\times&
\frac{\partial }{\partial \Delta_0} \mathcal{F}_{0,X^{\prime}Y^{\prime}}^{-1}(\tau_2-\tau_3)[{\chi}_{\mathrm{pair}}^{\mathrm{loc}}(i\nu_{n}=0)]^{-1}
\end{eqnarray}
where again $\delta \Gamma_{h_p}$, is $\delta \Gamma_{\Delta_0}$, with $\Delta_0$ replaced with $h_p$, and the derivative is evaluated at $h_p=0$, where $h_p$ is a source field that couples to the pairing operators.
We can relate our result to that using the Bethe-Salpeter equation within DCA and conclude that this additional contribution
is related to the particle particle bubble contribution which only arises if there are both special and normal $\tilde{{\bf q}}$'s.


\begin{thebibliography}{31}
\expandafter\ifx\csname natexlab\endcsname\relax\def\natexlab#1{#1}\fi
\expandafter\ifx\csname bibnamefont\endcsname\relax
  \def\bibnamefont#1{#1}\fi
\expandafter\ifx\csname bibfnamefont\endcsname\relax
  \def\bibfnamefont#1{#1}\fi
\expandafter\ifx\csname citenamefont\endcsname\relax
  \def\citenamefont#1{#1}\fi
\expandafter\ifx\csname url\endcsname\relax
  \def\url#1{\texttt{#1}}\fi
\expandafter\ifx\csname urlprefix\endcsname\relax\def\urlprefix{URL }\fi
\providecommand{\bibinfo}[2]{#2}
\providecommand{\eprint}[2][]{\url{#2}}

\bibitem{Mathur_Nature1998} 
N.\ D.\ Mathur, F.\ M.\ Grosche, S.\ R.\ Julian, I.\ R.\ Walker, D.\ M.\ Freye, R.\ K.\ W.\ Haselwimmer, and G.\ G.\ Lonzarich,
Nature {\bf 394}, 39 (1998).

\bibitem{Park_Nature2006} 
T.\ Park, F.\ Ronning, H.\ Q.\ Yuan, M.\ B.\ Salamon, R.\ Movshovich, J.\ L.\ Sarrao, and J.\ D.\ Thompson,
Nature {\bf 440}, 65 (2006).

\bibitem{Stockert_Natphys2012} 
O.\ Stockert, J.\ Arndt, E.\ Faulhaber,	 C.\ Geibel, H.\ S.\ Jeevan, S.\ Kirchner, M.\ Loewenhaupt, K.\ Schmalzl, W.\ Schmidt, Q.\ Si, and F.\ Steglich,
Nat. Phys. {\bf 7}, 119 (2011).

\bibitem{SiSteglich_Science2010}
Q. Si and F. Steglich, {\bf 329}, 1161 (2010).

\bibitem{Hertz} J.\ A.\ Hertz, Phys. Rev. B {\bf 14}, 1165 (1976).

\bibitem{Millis} A. J. Millis, Phys. Rev. B {\bf 48}, 7183  (1993).

\bibitem{Moriya}  T.\ Moriya
 \emph{Spin Fluctuations in Itinerant Electron Magnetism},
 {\bf 56},
 44--81,
 Springer (1985).


\bibitem{Si-2001}
Q.\ Si, S.\ Rabello, K.\ Ingersent, and J.\ L.\ Smith,
Nature {\bf 413}, 804 (2001).
  
\bibitem{Coleman-2001} 
P.\ Coleman, C.\ P\'epin, Q.\ Si, and R.\ Ramazashvili,
J. Phys.: Condens. Matter {\bf 13}, R723 (2001). 

\bibitem{Senthil-2004} T.\ Senthil, M.\ Vojta, and S.\ Sachdev,
Phys. Rev. B {\bf 69}, 035111 (2004).

\bibitem{Si_etal_JPSJ2013}
Q.\ Si, J.\ H.\ Pixley, E.\ Nica, S.\ J.\ Yamamoto, P.\ Goswami, R.\ Yu, S.\ Kirchner, J. Phys. Soc. Jpn. \textbf{83} 061005 (2014). 


\bibitem[{\citenamefont{Georges et~al.}(1996)\citenamefont{Georges, Kotliar,
  Krauth, and Rozenberg}}]{DMFT-review}
\bibinfo{author}{\bibfnamefont{A.}~\bibnamefont{Georges}},
  \bibinfo{author}{\bibfnamefont{G.}~\bibnamefont{Kotliar}},
  \bibinfo{author}{\bibfnamefont{W.}~\bibnamefont{Krauth}}, \bibnamefont{and}
  \bibinfo{author}{\bibfnamefont{M.~J.} \bibnamefont{Rozenberg}},
  \bibinfo{journal}{Rev. Mod. Phys.} \textbf{\bibinfo{volume}{68}},
  \bibinfo{pages}{13} (\bibinfo{year}{1996}).

\bibitem[{\citenamefont{Metzner and Vollhardt}(1989)}]{MetznerVollhardt}
\bibinfo{author}{\bibfnamefont{W.}~\bibnamefont{Metzner}} \bibnamefont{and}
  \bibinfo{author}{\bibfnamefont{D.}~\bibnamefont{Vollhardt}},
  \bibinfo{journal}{Physical review letters} \textbf{\bibinfo{volume}{62}},
  \bibinfo{pages}{324} (\bibinfo{year}{1989}).

\bibitem[{\citenamefont{Si and Smith}(1996)}]{Si-1996}
\bibinfo{author}{\bibfnamefont{Q.}~\bibnamefont{Si}} \bibnamefont{and}
  \bibinfo{author}{\bibfnamefont{J.~L.} \bibnamefont{Smith}},
  \bibinfo{journal}{Phys. Rev. Lett.} \textbf{\bibinfo{volume}{77}},
  \bibinfo{pages}{3391} (\bibinfo{year}{1996}).

\bibitem[{\citenamefont{Smith and Si}(2000)}]{Smith}
\bibinfo{author}{\bibfnamefont{J.~L.} \bibnamefont{Smith}} \bibnamefont{and}
  \bibinfo{author}{\bibfnamefont{Q.}~\bibnamefont{Si}}, \bibinfo{journal}{Phys.
  Rev. B} \textbf{\bibinfo{volume}{61}}, \bibinfo{pages}{5184}
  (\bibinfo{year}{2000}).

\bibitem[{\citenamefont{Chitra and Kotliar}(2000)}]{Chitra-2000}
\bibinfo{author}{\bibfnamefont{R.}~\bibnamefont{Chitra}} \bibnamefont{and}
  \bibinfo{author}{\bibfnamefont{G.}~\bibnamefont{Kotliar}},
  \bibinfo{journal}{Physical Review B} \textbf{\bibinfo{volume}{62}},
  \bibinfo{pages}{12715} (\bibinfo{year}{2000}).


\bibitem[{\citenamefont{Si et~al.}(2003)\citenamefont{Si, Rabello, Ingersent,
  and Smith}}]{Si-2003}
\bibinfo{author}{\bibfnamefont{Q.}~\bibnamefont{Si}},
  \bibinfo{author}{\bibfnamefont{S.}~\bibnamefont{Rabello}},
  \bibinfo{author}{\bibfnamefont{K.}~\bibnamefont{Ingersent}},
  \bibnamefont{and} \bibinfo{author}{\bibfnamefont{J.~L.} \bibnamefont{Smith}},
  \bibinfo{journal}{Physical Review B} \textbf{\bibinfo{volume}{68}},
  \bibinfo{pages}{115103} (\bibinfo{year}{2003}).

\bibitem[{\citenamefont{Zhu et~al.}(2003)\citenamefont{Zhu, Grempel, and
  Si}}]{Zhu.2003}
\bibinfo{author}{\bibfnamefont{J.-X.} \bibnamefont{Zhu}},
  \bibinfo{author}{\bibfnamefont{D.~R.} \bibnamefont{Grempel}},
  \bibnamefont{and} \bibinfo{author}{\bibfnamefont{Q.}~\bibnamefont{Si}},
  \bibinfo{journal}{Phys. Rev. Lett.} \textbf{\bibinfo{volume}{91}},
  \bibinfo{pages}{156404} (\bibinfo{year}{2003}).

\bibitem[{\citenamefont{Glossop and Ingersent}(2007)}]{Glossop.2007}
\bibinfo{author}{\bibfnamefont{M.~T.} \bibnamefont{Glossop}} \bibnamefont{and}
  \bibinfo{author}{\bibfnamefont{K.}~\bibnamefont{Ingersent}},
  \bibinfo{journal}{Physical review letters} \textbf{\bibinfo{volume}{99}},
  \bibinfo{pages}{227203} (\bibinfo{year}{2007}).

\bibitem[{\citenamefont{Zhu et~al.}(2007)\citenamefont{Zhu, Kirchner, Bulla,
  and Si}}]{Zhu.2007}
\bibinfo{author}{\bibfnamefont{J.-X.} \bibnamefont{Zhu}},
  \bibinfo{author}{\bibfnamefont{S.}~\bibnamefont{Kirchner}},
  \bibinfo{author}{\bibfnamefont{R.}~\bibnamefont{Bulla}}, \bibnamefont{and}
  \bibinfo{author}{\bibfnamefont{Q.}~\bibnamefont{Si}},
  \bibinfo{journal}{Physical review letters} \textbf{\bibinfo{volume}{99}},
  \bibinfo{pages}{227204} (\bibinfo{year}{2007}).

\bibitem[{\citenamefont{Zhu and Si}(2002)}]{Zhu-02}
\bibinfo{author}{\bibfnamefont{L.}~\bibnamefont{Zhu}} \bibnamefont{and}
  \bibinfo{author}{\bibfnamefont{Q.}~\bibnamefont{Si}},
  \bibinfo{journal}{Physical Review B} \textbf{\bibinfo{volume}{66}},
  \bibinfo{pages}{024426} (\bibinfo{year}{2002}).
  
    \bibitem{Aronson.95+Schroeder.00}
\bibinfo{author}{\bibnamefont{A.} \bibnamefont{Schr\"oder}},
\bibinfo{author}{\bibnamefont{G.} \bibnamefont{Aeppli}},
\bibinfo{author}{\bibnamefont{R.} \bibnamefont{Coldea}},
\bibinfo{author}{\bibnamefont{M.} \bibnamefont{Adams}},
\bibinfo{author}{\bibnamefont{O.} \bibnamefont{Stockert}},
\bibinfo{author}{\bibfnamefont{H. v.} \bibnamefont{L\"{o}hneysen}},
\bibinfo{author}{\bibnamefont{E.} \bibnamefont{Bucher}}
\bibinfo{author}{\bibnamefont{R.} \bibnamefont{Ramazashvili}} \bibnamefont{and}
\bibinfo{author}{\bibnamefont{P.} \bibnamefont{Coleman}},
\bibinfo{year}{2000}
  \bibinfo{journal}{Nature} \textbf{\bibinfo{volume}{407}},
  \bibinfo{pages}{351} (\bibinfo{year}{2000});
  \bibinfo{author}{\bibfnamefont{M. C.} \bibnamefont{Aronson}},
\bibinfo{author}{\bibfnamefont{R.} \bibnamefont{Osborn}},
\bibinfo{author}{\bibfnamefont{R. A.} \bibnamefont{Robinson}},
\bibinfo{author}{ \bibfnamefont{J. W.} \bibnamefont{Lynn}},
\bibinfo{author}{\bibfnamefont{R.} \bibnamefont{Chau}},
\bibinfo{author}{ \bibfnamefont{C. L.} \bibnamefont{Seaman}} \bibnamefont{and}
\bibinfo{author}{ \bibfnamefont{M. B.} \bibnamefont{Maple}},
  \bibinfo{journal}{Phys. Rev. Lett.} \textbf{\bibinfo{volume}{75}},
  \bibinfo{pages}{725} (\bibinfo{year}{1995}).


\bibitem{Paschen.00+Gegenwart.07+Friedemann.10}
\bibinfo{author}{\bibfnamefont{S.} \bibnamefont{Paschen}},
\bibinfo{author}{\bibfnamefont{T.} \bibnamefont{L\"uhmann}},
\bibinfo{author}{\bibfnamefont{S.} \bibnamefont{Wirth}},
\bibinfo{author}{\bibfnamefont{P.} \bibnamefont{Gegenwart}},
\bibinfo{author}{\bibfnamefont{O.} \bibnamefont{Trovarelli}},
\bibinfo{author}{\bibfnamefont{C.} \bibnamefont{Geibel}},
\bibinfo{author}{\bibfnamefont{F.} \bibnamefont{Steglich}},
\bibinfo{author}{\bibfnamefont{P.} \bibnamefont{Coleman}} \bibnamefont{and}
\bibinfo{author}{ \bibfnamefont{Q.} \bibnamefont{Si}}, 
 \bibinfo{journal}{Nature} \textbf{\bibinfo{volume}{432}},
\bibinfo{pages}{881} (\bibinfo{year}{2004});
P.\ Gegenwart, T.\ Westerkamp, C.\ Krellner, Y.\ Tokiwa, S.\ Paschen, C.\ Geibel, F.\ Steglich, E.\ Abrahams, Q.\ Si,
Science, \textbf{315}, 969 (2007);
\bibinfo{author}{\bibfnamefont{S.} \bibnamefont{Friedemann}},
\bibinfo{author}{ \bibfnamefont{N.} \bibnamefont{Oeschler}},
\bibinfo{author}{\bibfnamefont{S.} \bibnamefont{Wirth}},
\bibinfo{author}{\bibfnamefont{C.} \bibnamefont{Krellner}},
\bibinfo{author}{\bibfnamefont{C.} \bibnamefont{Geibel}},
\bibinfo{author}{ \bibfnamefont{F.} \bibnamefont{Steglich}},
\bibinfo{author}{\bibfnamefont{S.} \bibnamefont{Paschen}},
\bibinfo{author}{\bibfnamefont{S.} \bibnamefont{Kirchner}} \bibnamefont{and}
\bibinfo{author}{\bibfnamefont{Q.} \bibnamefont{Si}},
\bibinfo{journal}{Proc. Natl. Acad. Sci. USA}
\textbf{\bibinfo{volume}{107}}, \bibinfo{pages}{14547} (\bibinfo{year}{2010}).

\bibitem{Shishido}
 \bibinfo{author}{\bibfnamefont{H.}~\bibnamefont{Shishido}},
\bibinfo{author}{\bibfnamefont{R.}~\bibnamefont{Settai}},
\bibinfo{author}{\bibfnamefont{H.}~\bibnamefont{Harima}}, and
\bibinfo{author}{\bibfnamefont{Y.}~\bibnamefont{\={O}nuki}},
  \bibinfo{journal}{J.~Phys.~Soc.~Jpn.} \textbf{\bibinfo{volume}{74}},
  \bibinfo{pages}{1103} (\bibinfo{year}{2005}).
  
  \bibitem{SiPashchen13}
Q.\ Si and S.\ Paschen,
  Phys.\ Status Solidi B \textbf{250}, 425 (2013).

\bibitem[{\citenamefont{Maier et~al.}(2005)\citenamefont{Maier, Jarrell,
  Pruschke, and Hettler}}]{cluster-review}
\bibinfo{author}{\bibfnamefont{T.}~\bibnamefont{Maier}},
  \bibinfo{author}{\bibfnamefont{M.}~\bibnamefont{Jarrell}},
  \bibinfo{author}{\bibfnamefont{T.}~\bibnamefont{Pruschke}}, \bibnamefont{and}
  \bibinfo{author}{\bibfnamefont{M.~H.} \bibnamefont{Hettler}},
  \bibinfo{journal}{Rev. Mod. Phys.} \textbf{\bibinfo{volume}{77}},
  \bibinfo{pages}{1027} (\bibinfo{year}{2005}).

\bibitem[{\citenamefont{Kotliar et~al.}(2001)\citenamefont{Kotliar, Savrasov,
  P\'alsson, and Biroli}}]{cdmft}
\bibinfo{author}{\bibfnamefont{G.}~\bibnamefont{Kotliar}},
  \bibinfo{author}{\bibfnamefont{S.~Y.} \bibnamefont{Savrasov}},
  \bibinfo{author}{\bibfnamefont{G.}~\bibnamefont{P\'alsson}},
  \bibnamefont{and} \bibinfo{author}{\bibfnamefont{G.}~\bibnamefont{Biroli}},
  \bibinfo{journal}{Phys. Rev. Lett.} \textbf{\bibinfo{volume}{87}},
  \bibinfo{pages}{186401} (\bibinfo{year}{2001}).

\bibitem[{\citenamefont{Hettler et~al.}(1998)\citenamefont{Hettler,
  Tahvildar-Zadeh, Jarrell, Pruschke, and Krishnamurthy}}]{DCA}
\bibinfo{author}{\bibfnamefont{M.~H.} \bibnamefont{Hettler}},
  \bibinfo{author}{\bibfnamefont{A.~N.} \bibnamefont{Tahvildar-Zadeh}},
  \bibinfo{author}{\bibfnamefont{M.}~\bibnamefont{Jarrell}},
  \bibinfo{author}{\bibfnamefont{T.}~\bibnamefont{Pruschke}}, \bibnamefont{and}
  \bibinfo{author}{\bibfnamefont{H.~R.} \bibnamefont{Krishnamurthy}},
  \bibinfo{journal}{Phys. Rev. B} \textbf{\bibinfo{volume}{58}},
  \bibinfo{pages}{R7475} (\bibinfo{year}{1998}).

\bibitem[{\citenamefont{Potthoff et~al.}(2003)\citenamefont{Potthoff, Aichhorn,
  and Dahnken}}]{VCA}
\bibinfo{author}{\bibfnamefont{M.}~\bibnamefont{Potthoff}},
  \bibinfo{author}{\bibfnamefont{M.}~\bibnamefont{Aichhorn}}, \bibnamefont{and}
  \bibinfo{author}{\bibfnamefont{C.}~\bibnamefont{Dahnken}},
  \bibinfo{journal}{Phys. Rev. Lett.} \textbf{\bibinfo{volume}{91}},
  \bibinfo{pages}{206402} (\bibinfo{year}{2003}).

\bibitem[{\citenamefont{S\'en\'echal et~al.}(2000)\citenamefont{S\'en\'echal,
  Perez, and Pioro-Ladri\`ere}}]{CPT}
\bibinfo{author}{\bibfnamefont{D.}~\bibnamefont{S\'en\'echal}},
  \bibinfo{author}{\bibfnamefont{D.}~\bibnamefont{Perez}}, \bibnamefont{and}
  \bibinfo{author}{\bibfnamefont{M.}~\bibnamefont{Pioro-Ladri\`ere}},
  \bibinfo{journal}{Phys. Rev. Lett.} \textbf{\bibinfo{volume}{84}},
  \bibinfo{pages}{522} (\bibinfo{year}{2000}).

\bibitem[{\citenamefont{Haule and Kotliar}(2007)}]{Haule-2007}
\bibinfo{author}{\bibfnamefont{K.}~\bibnamefont{Haule}} \bibnamefont{and}
  \bibinfo{author}{\bibfnamefont{G.}~\bibnamefont{Kotliar}},
  \bibinfo{journal}{Phys. Rev. B} \textbf{\bibinfo{volume}{76}},
  \bibinfo{pages}{104509} (\bibinfo{year}{2007}).

\bibitem[{\citenamefont{Hettler et~al.}(2000)\citenamefont{Hettler, Mukherjee,
  Jarrell, and Krishnamurthy}}]{DCA2}
\bibinfo{author}{\bibfnamefont{M.~H.} \bibnamefont{Hettler}},
  \bibinfo{author}{\bibfnamefont{M.}~\bibnamefont{Mukherjee}},
  \bibinfo{author}{\bibfnamefont{M.}~\bibnamefont{Jarrell}}, \bibnamefont{and}
  \bibinfo{author}{\bibfnamefont{H.~R.} \bibnamefont{Krishnamurthy}},
  \bibinfo{journal}{Phys. Rev. B} \textbf{\bibinfo{volume}{61}},
  \bibinfo{pages}{12739} (\bibinfo{year}{2000}).


\bibitem[{\citenamefont{Sun and Kotliar}(2002)}]{Sun}
\bibinfo{author}{\bibfnamefont{P.}~\bibnamefont{Sun}} \bibnamefont{and}
  \bibinfo{author}{\bibfnamefont{G.}~\bibnamefont{Kotliar}},
  \bibinfo{journal}{Phys. Rev. B} \textbf{\bibinfo{volume}{66}},
  \bibinfo{pages}{085120} (\bibinfo{year}{2002}).
  
  \bibitem[{\citenamefont{Georges et~al.}(2001)\citenamefont{Georges,
  Siddharthan, and Florens}}]{RVB-DMFT}
\bibinfo{author}{\bibfnamefont{A.}~\bibnamefont{Georges}},
  \bibinfo{author}{\bibfnamefont{R.}~\bibnamefont{Siddharthan}},
  \bibnamefont{and} \bibinfo{author}{\bibfnamefont{S.}~\bibnamefont{Florens}},
  \bibinfo{journal}{Phys. Rev. Lett.} \textbf{\bibinfo{volume}{87}},
  \bibinfo{pages}{277203} (\bibinfo{year}{2001}).

  
  
\bibitem[{\citenamefont{Metzner}(1991)}]{Metzner}
\bibinfo{author}{\bibfnamefont{W.}~\bibnamefont{Metzner}},
  \bibinfo{journal}{Phys. Rev. B} \textbf{\bibinfo{volume}{43}},
  \bibinfo{pages}{8549} (\bibinfo{year}{1991}).


\bibitem[{\citenamefont{Biroli and Kotliar}(2002)}]{biroli2002}
\bibinfo{author}{\bibfnamefont{G.}~\bibnamefont{Biroli}} \bibnamefont{and}
  \bibinfo{author}{\bibfnamefont{G.}~\bibnamefont{Kotliar}},
  \bibinfo{journal}{Phys. Rev. B} \textbf{\bibinfo{volume}{65}},
  \bibinfo{pages}{155112} (\bibinfo{year}{2002}).

\bibitem[{\citenamefont{Si et~al.}(2005)\citenamefont{Si, Zhu, and
  Grempel}}]{Si-2005}
\bibinfo{author}{\bibfnamefont{Q.}~\bibnamefont{Si}},
  \bibinfo{author}{\bibfnamefont{J.-X.} \bibnamefont{Zhu}}, \bibnamefont{and}
  \bibinfo{author}{\bibfnamefont{D.}~\bibnamefont{Grempel}},
  \bibinfo{journal}{Journal of Physics: Cond. Matt.}
  \textbf{\bibinfo{volume}{17}}, \bibinfo{pages}{R1025} (\bibinfo{year}{2005}).

\bibitem[{\citenamefont{Baskaran et~al.}(1987)\citenamefont{Baskaran, Zou, and
  Anderson}}]{baskaran1987}
\bibinfo{author}{\bibfnamefont{G.}~\bibnamefont{Baskaran}},
  \bibinfo{author}{\bibfnamefont{Z.}~\bibnamefont{Zou}}, \bibnamefont{and}
  \bibinfo{author}{\bibfnamefont{P.}~\bibnamefont{Anderson}},
  \bibinfo{journal}{Solid State Communications} \textbf{\bibinfo{volume}{63}},
  \bibinfo{pages}{973} (\bibinfo{year}{1987}).

\bibitem[{\citenamefont{Mineev et~al.}(1999)\citenamefont{Mineev, Samokhin, and
  Landau}}]{Mineev-book}
\bibinfo{author}{\bibfnamefont{V.~P.} \bibnamefont{Mineev}},
  \bibinfo{author}{\bibfnamefont{K.}~\bibnamefont{Samokhin}}, \bibnamefont{and}
  \bibinfo{author}{\bibfnamefont{L.~D.} \bibnamefont{Landau}},
  \emph{\bibinfo{title}{Introduction to unconventional superconductivity}}
  (\bibinfo{publisher}{CRC Press}, \bibinfo{year}{1999}).

\bibitem[{\citenamefont{Pixley et~al.}(2013)\citenamefont{Pixley, Deng,
  Ingersent, and Si}}]{Pixley-2013}
\bibinfo{author}{\bibfnamefont{J.~H.}~\bibnamefont{Pixley}},
  \bibinfo{author}{\bibfnamefont{L.}~\bibnamefont{Deng}},
  \bibinfo{author}{\bibfnamefont{K.}~\bibnamefont{Ingersent}},
  \bibnamefont{and} \bibinfo{author}{\bibfnamefont{Q.}~\bibnamefont{Si}},
  \bibinfo{journal}{arXiv:1308.0839}  (\bibinfo{year}{2013}).

\bibitem[{\citenamefont{Freedman et~al.}(2005)\citenamefont{Freedman, Nayak,
  and Shtengel}}]{Freedman-2005}
\bibinfo{author}{\bibfnamefont{M.}~\bibnamefont{Freedman}},
  \bibinfo{author}{\bibfnamefont{C.}~\bibnamefont{Nayak}}, \bibnamefont{and}
  \bibinfo{author}{\bibfnamefont{K.}~\bibnamefont{Shtengel}},
  \bibinfo{journal}{Physical review letters} \textbf{\bibinfo{volume}{94}},
  \bibinfo{pages}{066401} (\bibinfo{year}{2005}).

\bibitem[{\citenamefont{Witczak-Krempa
  et~al.}(2014)\citenamefont{Witczak-Krempa, Chen, Kim, and
  Balents}}]{Witczak-2014}
\bibinfo{author}{\bibfnamefont{W.}~\bibnamefont{Witczak-Krempa}},
  \bibinfo{author}{\bibfnamefont{G.}~\bibnamefont{Chen}},
  \bibinfo{author}{\bibfnamefont{Y.~B.} \bibnamefont{Kim}}, \bibnamefont{and}
  \bibinfo{author}{\bibfnamefont{L.}~\bibnamefont{Balents}},
  \bibinfo{journal}{Annual Review of Condensed Matter Physics}
  \textbf{\bibinfo{volume}{5}}, \bibinfo{pages}{57} (\bibinfo{year}{2014}).

\bibitem[{\citenamefont{Farrell and Pereg-Barnea}(2014)}]{Farrel-2014}
\bibinfo{author}{\bibfnamefont{A.}~\bibnamefont{Farrell}} \bibnamefont{and}
  \bibinfo{author}{\bibfnamefont{T.}~\bibnamefont{Pereg-Barnea}},
  \bibinfo{journal}{Phys. Rev. B} \textbf{\bibinfo{volume}{89}},
  \bibinfo{pages}{035112} (\bibinfo{year}{2014}).


\bibitem[{\citenamefont{Werner and Millis}(2007)}]{Werner-2007}
\bibinfo{author}{\bibfnamefont{P.}~\bibnamefont{Werner}} \bibnamefont{and}
  \bibinfo{author}{\bibfnamefont{A.~J.} \bibnamefont{Millis}},
  \bibinfo{journal}{Phys. Rev. Lett.} \textbf{\bibinfo{volume}{99}},
  \bibinfo{pages}{146404} (\bibinfo{year}{2007}).

\bibitem[{\citenamefont{Werner and Millis}(2010)}]{Werner-2010}
\bibinfo{author}{\bibfnamefont{P.}~\bibnamefont{Werner}} \bibnamefont{and}
  \bibinfo{author}{\bibfnamefont{A.~J.} \bibnamefont{Millis}},
  \bibinfo{journal}{Phys. Rev. Lett.} \textbf{\bibinfo{volume}{104}},
  \bibinfo{pages}{146401} (\bibinfo{year}{2010}).

\bibitem[{\citenamefont{Pixley et~al.}(2011)\citenamefont{Pixley, Kirchner,
  Glossop, and Si}}]{Pixley-2010}
\bibinfo{author}{\bibfnamefont{J.~H.} \bibnamefont{Pixley}},
  \bibinfo{author}{\bibfnamefont{S.}~\bibnamefont{Kirchner}},
  \bibinfo{author}{\bibfnamefont{M.~T.} \bibnamefont{Glossop}},
  \bibnamefont{and} \bibinfo{author}{\bibfnamefont{Q.}~\bibnamefont{Si}}, 
  \emph{\bibinfo{booktitle}{Journal of Physics: Conference Series}}
  (\bibinfo{organization}{IOP Publishing}, \bibinfo{year}{2011}), vol.
  \bibinfo{volume}{273}, p. \bibinfo{pages}{012050}.
  
  \bibitem[{\citenamefont{Pixley}(2013)}]{Pixley2-2013}
\bibinfo{author}{\bibfnamefont{J.~H.}~\bibnamefont{Pixley}} 
\bibinfo{author}{\bibfnamefont{S.}~\bibnamefont{Kirchner}}
\bibinfo{author}{\bibfnamefont{K.}~\bibnamefont{Ingersent}}  
\bibnamefont{and}
  \bibinfo{author}{\bibfnamefont{Q.} \bibnamefont{Si}},
  \bibinfo{journal}{Phys. Rev. B} \textbf{\bibinfo{volume}{88}},
  \bibinfo{pages}{245111} (\bibinfo{year}{2013}).
 


\bibitem[{\citenamefont{Otsuki}(2013)}]{Otsuki-2013}
\bibinfo{author}{\bibfnamefont{J.}~\bibnamefont{Otsuki}},
  \bibinfo{journal}{Phys. Rev. B} \textbf{\bibinfo{volume}{87}},
  \bibinfo{pages}{125102} (\bibinfo{year}{2013}).

\bibitem[{\citenamefont{\ifmmode~\check{Z}\else
  \v{Z}\fi{}itko}(2009)}]{Zitko-2009}
\bibinfo{author}{\bibfnamefont{R.}~\bibnamefont{\ifmmode~\check{Z}\else
  \v{Z}\fi{}itko}}, \bibinfo{journal}{Phys. Rev. B}
  \textbf{\bibinfo{volume}{80}}, \bibinfo{pages}{125125}
  (\bibinfo{year}{2009}).
  
  \bibitem[{\citenamefont{Baym and Kadanoff}(1961)}]{Baym-1961}
\bibinfo{author}{\bibfnamefont{G.}~\bibnamefont{Baym}} \bibnamefont{and}
  \bibinfo{author}{\bibfnamefont{L.~P.} \bibnamefont{Kadanoff}},
  \bibinfo{journal}{Phys. Rev.} \textbf{\bibinfo{volume}{124}},
  \bibinfo{pages}{287} (\bibinfo{year}{1961}).



  
\bibitem{Leo-2008}  
  L.\ D.\ Leo, M.\ Civelli, and G.\ Kotliar,
  Phys. Rev. B {\bf 77}, 075107 (2008);
  Phys. Rev. Lett. {\bf 101}, 256404 (2008).


\end{thebibliography}
\end{document}